\documentclass[prd,aps,preprintnumbers,nofootinbib,preprint,floatfix,superscriptaddress]{revtex4}
\pdfoutput=1
\usepackage{amsmath,amssymb}
\usepackage{cancel}

\usepackage{hyperref}
\usepackage{natbib}
\usepackage{anysize}
\usepackage{graphicx}
\usepackage{appendix}
\usepackage{slashed}

\marginsize{3cm}{3cm}{3cm}{3cm}

\begin{document}

\title{\huge $h\rightarrow\gamma\gamma$ excess and Dark Matter\\ from Composite Higgs Models\vspace{2cm}}

\author{\Large Mikael Chala}

\affiliation{
CAFPE and Departamento de F\'{\i}sica Te\'orica y del Cosmos, \\
Universidad de Granada, E-18071 Granada, Spain\vspace{1cm}
}

\begin{abstract}
Composite Higgs Models are very appealing candidates for a natural realization of electroweak symmetry breaking. Non minimal models could explain the recent Higgs data from ATLAS, CMS and Tevatron experiments, including the excess in the amount of diphoton events, as well as provide a natural dark matter candidate. In this article, we study a Composite Higgs model based on the coset $SO(7)/G2$. In addition to the Higgs doublet, one $SU(2)_L$ singlet of electric charge one, $\kappa^\pm$, as well as one singlet $\eta$ of the whole Standard Model group arise as pseudo-Goldstone bosons. $\kappa^\pm$ and $\eta$ can be responsible of the diphoton excess and dark matter respectively.

\end{abstract}

\maketitle


\section{Introduction}

Composite Higgs Models (CHM)\cite{Kaplan:1983fs,Kaplan:1983sm,Dimopoulos:1981xc} provide a compelling solution to the hierarchy problem. In these models, the Higgs boson arises as a bound state of a new strongly interacting sector with a global symmetry group $G$ spontaneously broken to $H\subset G$. Therefore, its mass is protected by its finite size, and it becomes naturally light ---as the ATLAS, CMS and Tevatron experiments have recently revealed\cite{TEVNPH:2012ab,ATLAS-CONF-2012-019,CMS-PAS-HIG-12-008}--- due to its pseudo-Nambu Goldstone (pNGB) nature. In the Minimal Composite Higgs Model (MCHM)\cite{Agashe:2004rs,Contino:2006qr}, this symmetry breaking pattern is achieved by the coset $SO(5)/SO(4)$. In the so-called MCHM5, the SM fermions mix with resonances of the strong sector transforming in the $\mathbf{5}$ representation of $SO(5)$. The MCHM5, however, can accommodate neither a Dark Matter (DM) candidate nor solution to the recent diphoton excess. Therefore, other non-minimal CHMs have been considered in the literature\cite{Gripaios:2009pe,Mrazek:2011iu,Redi:2012ha,Bertuzzo:2012ya,Frigerio:2012uc}, which give very interesting new signatures at the LHC and DM searches. In fact, regarding the diphoton discrepancy, many alternatives to the SM scalar sector have been proposed in order to explain this possible excess\cite{Djouadi:1998az,Petriello:2002uu,Han:2003gf,Chen:2006cs,Dermisek:2007fi,Low:2009nj,Low:2009di,Cacciapaglia:2009ky,Casagrande:2010si,Cheung:2011nv,Carena:2011aa,Cao:2011pg,Batell:2011pz,Arvanitaki:2011ck,Barger:2012hv,ArkaniHamed:2012kq,Arhrib:2012yv,Alves:2011kc,Lee:2012wz,Kearney:2012zi,Kanemura:2012rs,oglekar:2012vc,Dorsner:2012pp,Almeida:2012bq,Draper:2012xt,Akeroyd:2012ms,Dawson:2012di,Christensen:2012ei,Carena:2012xa,Delgado:2012sm,Chun:2012jw}.

Here we present a new CHM based on the symmetry breaking pattern of $SO(7)$ to $G2$. In this case, an uncolored $SU(2)_L$ singlet charged scalar, $\kappa^\pm$, as well as a neutral singlet scalar, $\eta$, appear in the spectrum in addition to the SM Higgs doublet $H$. As we show below, $\kappa^\pm$ and $\eta$ can reproduce the observed deviation in $\gamma\gamma$ events and DM, respectively, in a natural way. The stability of the latter is guaranteed by a $\eta\rightarrow -\eta$ symmetry. This symmetry is preserved by a particular embedding of the elementary SM sector into spinorial $\mathbf{8}$ representations of $SO(7)$. The absence of anomalies in this group, which therefore can not break this parity symmetry, makes $\eta$ a very natural candidate for DM. The excess in $\gamma\gamma$ is also very interesting, since it could provide a good hint of a larger scalar sector (possibly composite) to be probed with the near future LHC data. Hints of the composite nature of this sector can be also indirectly looked for in Higgs production in association with a $t\bar{t}$ pair \cite{Carmona:2012my,Vignaroli:2012nf} already in the current LHC run. After the longer LHC run with $\sqrt{s} = 14\,\text{TeV}$ of center of mass energy, this composite nature could be probed through the direct production of new resonances \cite{Carmona:2012my,Barcelo:2011wu,Bini:2011zb,Vignaroli:2012nf,Brooijmans:2012yi} or even through the pair production of Higgs bosons\cite{Contino:2010mh,Grober:2010yv,Contino:2012xk,Gillioz:2012se}.

This article is structured as follows. In Section \ref{model} we introduce the group structure of the model and obtain the two-derivative scalar interactions described by the non-linear sigma model lagrangian. We also discuss the embedding of the SM fermions into representations of the whole group and construct the lagrangian quadratic in the fermion fields. In Section \ref{colemanp}, we discuss the Coleman-Weinberg effective potential for the scalars generated through loops of fermions. In Section \ref{couplings}, we calculate the couplings of $h, \eta$ and $\kappa^\pm$ to the SM fermions and gauge bosons, and compare them with other CHMs. In Section \ref{phenomenology} we discuss some phenomenological implications of the new states. This includes a study of the $h\rightarrow \gamma \gamma$ process, $\eta$ as a natural dark matter candidate and prospects for production of the new scalars at the LHC. We conclude with a summary discussion in Section \ref{conclusions}.

\section{$SO(7)/G2$ Composite Higgs Model}\label{model}

\begin{figure}
\includegraphics[width=\columnwidth]{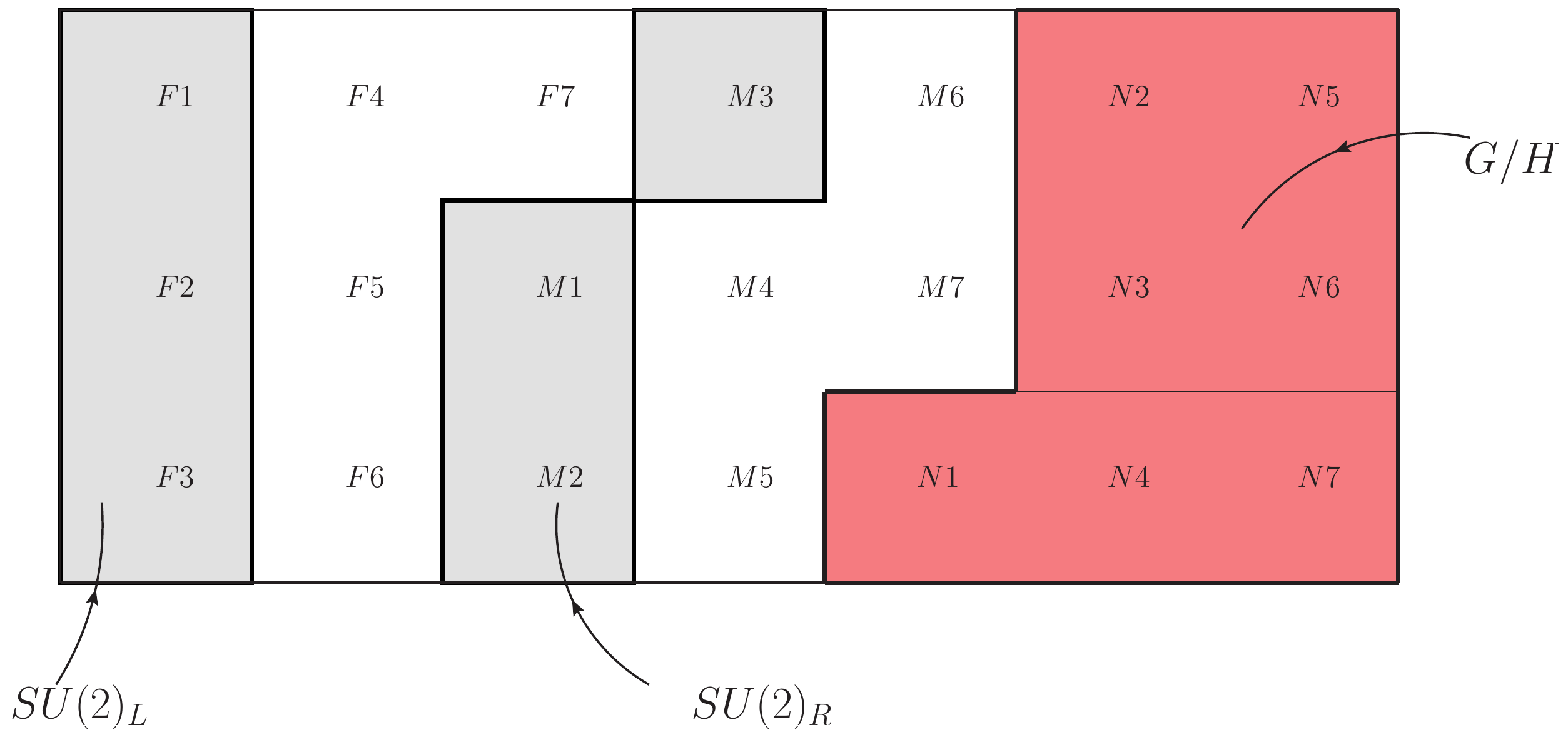}\hspace{-1.5cm}
\caption{Pictorial representation of the Lie algebra of $SO(7)$, and the embedding of its important subgroups $SU(2)_L\times SU(2)_R\subset G2$.}
\end{figure}\label{algebra}

The model is based on the symmetry breaking pattern $SO(7)/G2$, which can be achieved through the vev of a field $\Phi$ transforming in the spinor representation $\mathbf{8}$ of $SO(7)$. The 21 generators $J_{mn} = -J_{nm}$ of $SO(7)$ in this representation can be constructed out of the gamma matrices $\gamma_i$ of Appendix \ref{embeding} in the following way:
\begin{equation}
J_{mn} = -\frac{1}{4}\left[\gamma_m,\gamma_n\right].
\end{equation}
The group $G2$ can be then regarded as the set of elements of $SO(7)$ that leave the vacuum
\begin{equation}
v_0^T = (0,1,0,1,0,-1,0,-1)
\end{equation}
 invariant\cite{citeulike:8040542,Evans:1994np,Gunaydin:1995as}. The generators correspond to the sets $F_i$ and $M_i$ of Figure 1 (see Appendix \ref{embeding} for the explicit expressions). The rest of the generators, the set $N_i$, generate the coset manifold. They transform in the $\mathbf{7}$ representation of $G2$, and decompose under the subgroup $SU(2)_L\times SU(2)_R$ into $(\mathbf{2},\mathbf{2})\oplus(\mathbf{1},\mathbf{3})$. Some of the relevant commutation relations are
\begin{align}\label{com1}
 \left[F_i,F_j\right] &= i\epsilon^{ijk} F_k, \,\, [M_i,M_j] = \frac{i}{\sqrt{3}}\epsilon^{ijk} M_k, \,\, [F_i,M_j] = 0, \,\, [F_i, N_j] = 0,\\\nonumber
 [M_3,N_3] &= 0, \qquad \,\, [M_3,N^{\pm}] = \mp N^{\pm}, \qquad N^{\pm} = N_1\pm iN_2,
\end{align}

where $j=1,2,3$. All the $SO(7)$ generators of Figure 1 are normalized according to $\text{Tr}\left[ \text{T}_i \text{T}_j\right] = \delta_{ij}$ (note that the $SU(2)_R$ group is generated from $\sqrt{3}M_i$ rather than $M_i$ alone). From equation (\ref{com1}), we explicitly see how $N_{1,2,3}$ are not charged under $SU(2)_L$ and we get their hypercharges. In fact, these generators transform in the $(\mathbf{1},\mathbf{3})$ representation mentioned above (and so do the corresponding Goldstone bosons), while the rest of them live in the $(\mathbf{2},\mathbf{2})$, giving rise to the Higgs doublet. Thus, the pNGB spectrum is composed of the Higgs doublet $H$, a neutral scalar $\eta$ and a singly charged scalar $\kappa^\pm$.

\subsection{Scalar Sector}\label{ssector}
The scalar sector lagrangian is described by a non-linear sigma model over $SO(7)/G2$, with lagrangian 
\begin{equation}\label{sigmamodel}
\mathcal{L} = \frac{f^2}{6}\partial_\mu \Sigma^\dagger \partial^\mu \Sigma, 
\end{equation}
where
\begin{equation}\label{sigma}
\Sigma = e^{-i\sqrt{2}\Pi(x)/f}\Sigma_0, \qquad \Pi(x) = \Pi^i(x) N_i, \qquad \Sigma_0 = \langle\Phi\rangle = v_0.
\end{equation}
Since the $N_i$ coset generators are hermitian, $\Pi(x) = \Pi^\dagger(x)$. Therefore, if we expand the lagrangian up to $\mathcal{O}(1/f^4)$ we find
\begin{equation}
\mathcal{L} = \frac{f^2}{6}|\partial_\mu \Sigma|^2 = \frac{1}{3}\partial_\mu(\Pi\Sigma_0)^\dagger \partial^\mu(\Pi\Sigma_0)+\frac{1}{6f^2}\partial_\mu(\Pi^2\Sigma_0)^\dagger\partial^\mu(\Pi^2\Sigma_0)+\mathcal{O}(1/f^4).
\end{equation}
Using the explicit expression of $\Pi$ in Appendix \ref{chorizo}, we can write the lagrangian in terms of charge eigenstate fields. It reads, in the unitary gauge:
\begin{align}\label{sigmal}
\mathcal{L} =& \,\mathcal{K} + \frac{3}{8f^2}(\partial_\mu(H^\dagger H))^2+\frac{3}{8f^2}\eta^2(\partial_\mu\eta)^2+\frac{3}{8f^2}\partial_\mu (H^\dagger H)\partial^\mu\eta^2\\\nonumber
&+\frac{3}{8f^2}(\partial_\mu (\kappa^+\kappa^-))^2+\frac{3}{4f^2}\partial_\mu (H^\dagger H)\partial^\mu(\kappa^+\kappa^-) + \frac{3}{8f^2}\partial_\mu \eta^2\partial^\mu(\kappa^+ \kappa^-),
\end{align}
where $\mathcal{K}$ stands for the canonically normalized kinetic terms. We have defined $H^T=[(h_1+ih_2)/\sqrt{2},(h_3+ih_4)/\sqrt{2}]$ and $\kappa^\pm = (k_1\pm ik_2)/\sqrt{2}$.

\subsection{Fermion Spectrum}
In order to construct the effective lagrangian for the fermions, we should extend the symmetry group to $SO(7)\times U(1)_X$ and embed the SM fermions in multiplets of this group\cite{Agashe:2004rs}, with the proper $X$ charge. Two appropriate representations of the whole group $SO(7)$ are the fundamental $\mathbf{7}$ and the spinorial $\mathbf{8}$ representations. Under the unbroken subgroup $G2$, the first one remains as $\mathbf{7}$ while the second decomposes as $\mathbf{1}+\mathbf{7}$. We work in the latter scenario because, as we will see, the presence of a whole $G2$ singlet will be necessary to give a mass to the top quark. Under the custodial symmetry group $SU(2)_L\times SU(2)_R$, the $\mathbf{8}$ decomposes as $(\mathbf{1},\mathbf{1})+(\mathbf{2},\mathbf{2})+(\mathbf{1},\mathbf{3})$. The SM fermions will mix, therefore, with multiplets of $SO(7)$ of charge $\mathbf{8}_{2/3}$ and $\mathbf{8}_{-1/3}$. We pictorially\footnote{Pictorially in the sense that we are not writing the generators of Appendix \ref{embeding} in the canonical base of equation (\ref{pic}). Thus, for instance, in the base used for the generators, the custodial $(1,1)$ of the whole 8-dimensional space is not $(0,0,0,0,0,0,0,1)^T$ but rather $(0,1,0,1,0,-1,0,-1)^T.$ Equations (\ref{qul}), (\ref{terre}) and (\ref{berre}) should then be clear in light of this consideration.} represent the $\mathbf{8}$ as
\begin{equation}\label{pic}
\mathbf{8}_{2/3} = \left[\begin{array}{c}
                          (2,2) = (q,Q)\\
			  (1,3) = \left(\begin{array}{c}
				    X\\
				    t'\\
				    b
				   \end{array}\right)\\
			  (1,1) = t
                         \end{array}\right], \qquad \mathbf{8}_{-1/3} = \left[\begin{array}{c}
                          (2,2) = (Q',q')\\
			  (1,3) = \left(\begin{array}{c}
				    t''\\
				    b''\\
				    Y
				   \end{array}\right)\\
			  (1,1) = b'
                         \end{array}\right],
\end{equation}
%
%
where the decomposition into irreps of $SU(2)_L\times SU(2)_R$ is manifest. Let us discuss how the SM fields should be divided among the different entries. In order to give a mass $m_b$ to the bottom quark, both the left-handed components of the quarks along $q_L'$ as well as the component of the bottom quark along $b_R'$ should be different from zero. The component along $q_L'$, however, has to be small enough to protect the $Zb_L\bar{b}_L$ coupling\cite{Agashe:2006at}, and then the component of $b$ along $b_R'$ should be near one to naturally get a non-negligible $m_b$. Thus, the component of the bottom quark along $b_R$ can no longer be large. This, however, will only affect the $\kappa^\pm$ decay width. $b_R''$ can be fixed to zero without any conflict. On the other hand, the component of the top quark along $t_R$ should be nearly one to allow a naturally large top mass, making the component along $t_R'$ rather small. In fact, if this component is non-vanishing the  $\eta\rightarrow -\eta$ parity symmetry would be explicitly broken.

Let us focus on the top sector, which naturally contains the largest couplings. We choose a prescription consisting of two $\mathbf{8}_{2/3}$ fields, $Q_L$ and $T_R$. The SM doublets can be embedded in the $(\mathbf{2},\mathbf{2})$ of $Q_L$, where the $Zb_L\bar{b}_L$ coupling becomes protected as mentioned above:
\begin{equation}\label{qul}
Q_L = \frac{1}{\sqrt{8}}(it_L-b_L, t_L-ib_L,-it_L-b_L,t_L+ib_L,it_L+b_L,b_L-it_L,t_L-ib_L)^T.
\end{equation}
The $T_R$ field can contain the $t_R$ singlet in both the $(\mathbf{1},\mathbf{1})$ and the neutral part of $(\mathbf{1},\mathbf{3})$, proportional to $\cos{\theta}$ and $\sin{\theta}$ respectively, and also a small component of the $b_R$ field in the same $\mathbf{(1,3)}$:
\begin{equation}\label{terre}
 T_R = \frac{1}{2}(s_{\theta}t_R,c_{\theta}t_R,-s_{\theta}t_R,c_{\theta}t_R,-s_{\theta}t_R,-c_{\theta}t_R,s_{\theta}t_R,-c_{\theta}t_R)^T+\epsilon B_R^T ,
\end{equation}
with
\begin{equation}\label{berre}
B_R = \frac{1}{\sqrt{8}}(ib_R,b_R,ib_R,-b_R,ib_R,b_R,ib_R,-b_R)^T.
\end{equation}
%
%
%
Although the embedding of $b_R$ in $B_R$ does not give a mass to the bottom quark, $\epsilon$ has to be different from zero. Otherwise, $\kappa^\pm$ appears always in pairs and then becomes stable, giving rise to undesirable consequences\cite{DeRújula1990173,PhysRevD.41.2388,PhysRevLett.65.957,Gould:1989gw}. The hypercharge $Y$ of the different elementary fields is $Y = T_R^3+Q_X$, where $T_R^3$ refers to the third generator of $SU(2)_R$. Note that, as we will see, whenever $\sin{\theta}$ is different from zero, a trilinear coupling for $\eta$ is generated, allowing it to decay into pairs of fermions. So, if we want $\eta$ to be a DM candidate, we should set\footnote{This choice means that $t_R$ can not mix with the corresponding heavy resonance. We are thus recovering a parity symmetry (under which both $\eta$ and this heavy resonance are odd, while the rest of the particles remain even) that forbids this mixing to appear at the loop level.} $\theta = 0$. The most general $SO(7)\times U(1)_X$ invariant lagrangian of order two in the fields reads:
%
%
\begin{align}\label{order2}
\mathcal{L}_\text{eff} &=  \bar{T}_R \cancel{p} \left( \Pi_{t_R}^0 + \Pi^1_{t_R}\Sigma^T\Sigma\right) T_R + \bar{Q}_L\cancel{p} \left( \Pi_{qL}^0 +\Pi^1_{q_L}\Sigma^T\Sigma\right) Q_L \\\nonumber
 &+\left[f  M_t\, \bar{Q}_L\Sigma^T\Sigma T_R + h.c.\right].
\end{align}
After expanding $\Sigma$ up to $1/f^2$ we get, in the unitary gauge, the following effective lagrangian for the quarks:
\begin{align}\label{yukawa_end}
\nonumber&\mathcal{L}_\text{eff} = \bar{t}_R\cancel{p}\left(\Pi_{t_R}^0 + 3\Pi^1_{t_R}\left[\frac{4}{3}c_{\theta}^2+s_{\theta}^2\frac{\eta^2}{f^2}-\frac{c_{\theta}^2}{f^2}\left(h^2+\eta^2+2\kappa^+\kappa^-\right)+\frac{4\sqrt{3}}{3}c_{\theta}s_{\theta}\frac{\eta}{f}\right]\right)t_R\\\nonumber 
&+ \bar{t}_L\cancel{p}\left(\Pi_{q_L}^0 + \Pi_{q_L}^1 \frac{3h^2}{2f^2}\right)t_L + \Pi_{q_L}^0\bar{b}_L\cancel{p}b_L + \epsilon^2\bar{b}_R\cancel{p}\left(\Pi_{t_R}^0+3\frac{\Pi_{t_R}^1}{f^2}\kappa^+\kappa^-\right)b_R \\\nonumber
&+\bigg\{\Pi^1_{t_R}\epsilon c_\theta\bar{t}_R\cancel{p}\frac{\kappa^+}{f}\left(2\sqrt{3}+3\tan{\theta}\frac{\eta}{f}\right)b_R+\frac{3\sqrt{2}}{2}\epsilon M_t\bar{t}_Lb_R h \frac{\kappa^+}{f}\\
& + \sqrt{6}M_tc_{\theta}\,\bar{t}_Lt_Rh\left[1-\frac{3}{8f^2}\left(h^2+\eta^2+2\kappa^+\kappa^-\right)+\frac{3}{2\sqrt{3}}\tan{\theta}\,\frac{\eta}{f}\right]+h.c.\bigg\},
\end{align}
where $c_\theta\equiv \cos{\theta}$ and $s_\theta\equiv\sin{\theta}$. After EWSB, we obtain trilinear couplings of $\kappa^\pm$ to the fermions whenever $\epsilon$ is different from zero, that allows $\kappa^\pm$ to decay into SM particles. From here on we will consider $\cos{\theta} = 1$, since otherwise we would break the $\eta\rightarrow -\eta$ symmetry.

\section{Effective potential}\label{colemanp}

%
%
The embedding of the SM fermions into representations of the full strong symmetry group breaks explicitly the $SO(7)\times U(1)_X$ symmetry. Therefore, loops of fermions will generate a Coleman-Weinberg effective potential\cite{PhysRevD.7.1888} for the scalars. The main contribution comes from the top, while gauge contributions aligned with zero vev\cite{PhysRevLett.51.2351} are relevant for detailed calculations. The one loop potential $V(h,\eta,\kappa)$ is then given by the expression\cite{PhysRevD.9.1686}
\begin{equation}
V(h,\eta,\kappa) = -2N_c\int \frac{d^4p}{(2\pi)^4} \log{\left(\text{det}\frac{\partial^2\mathcal{L}}{\partial\bar{\psi}_i\psi_j}\right)}
\end{equation}
where $\psi_i$ can be either $t_L$ or $t_R$. Using the effective lagrangian in equation (\ref{yukawa_end}), we obtain, in euclidean space,
\begin{equation}
V(h,\eta,\kappa) = -2N_c\int\frac{d^4p}{(2\pi)^2}\log{\left(p^2\Pi_L\Pi_R+|\Pi_{LR}|^2\right)}
\end{equation}
where
\begin{align}
&\Pi_L = \Pi_{q_L}^0+\frac{3}{2f^2}\Pi_{q_L}^1h^2, \\\nonumber
& \Pi_R = \Pi_{t_R}^0 + 3\Pi^1_{t_R}\left(\frac{4}{3}-\frac{h^2+\eta^2+2\kappa^+\kappa^-}{f^2}\right),\\\nonumber
&\Pi_{LR} = \sqrt{6}M_t h \left(1-\frac{3}{8f^2}\left(h^2+\eta^2+2\kappa^+\kappa^-\right)\right).
\end{align}
Expanding the logarithms up to quartic terms, we get the following potential for the scalars:
\begin{align}\label{potencial}
V(h,\eta,\kappa) =& -\frac{\mu^2_h}{2}h^2+\frac{\lambda_h}{4}h^4+\frac{\mu^2_\eta}{2}\eta^2+\frac{\lambda_\eta}{4}\eta^4+\mu^2_{\kappa}\kappa^+\kappa^- + \lambda_\kappa(\kappa^+\kappa^-)^2\\\nonumber
+&\frac{\lambda_{h\eta}}{2}h^2\eta^2+\lambda_{h\kappa}h^2\kappa^+\kappa^-+\lambda_{\eta\kappa}\eta^2\kappa^+\kappa^-.
\end{align}
The $\mu$ and $\lambda$ parameters can be written in terms of the form factors, which can be explicitly calculated in extra dimensions theories\cite{Agashe:2004rs} or by means of Weinberg sum rules\cite{PhysRevLett.18.507,Marzocca:2012zn,Pomarol:2012qf} in the large $N_c$ limit. Otherwise, they are free parameters to be constrained by the experiments. We have, in units of $f$:
\begin{align}\label{formfactors}
-\frac{\mu_h^2}{2} &= -2N_c\int \left(\frac{3}{2}\frac{\Pi^1_{q_L}}{\Pi^0_{q_L}}-\frac{3\Pi^1_{t_R}}{\Pi^0_{t_R}+4\Pi^1_{t_R}}+\frac{6|M_t|^2}{p^2\Pi^0_{q_L}(\Pi^0_{t_R}+4\Pi^1_{t_R})}\right), \\\nonumber
\frac{\mu^2_\eta}{2}& = \frac{\mu^2_\kappa}{2} = -2N_c \int \left(-\frac{3\Pi^1_{t_R}}{\Pi^0_{t_R}+4\Pi^1_{t_R}}\right), \\\nonumber
\frac{\lambda_\eta}{4} & = \frac{\lambda_{\kappa}}{4} = \frac{\lambda_{\eta\kappa}}{2} = -2N_c \int -\frac{1}{2}\left(\frac{3\Pi^1_{t_R}}{\Pi^0_{t_R}+4\Pi^1_{t_R}}\right)^2, \\\nonumber
\frac{\lambda_{h\eta}}{2} &= \frac{\lambda_{h\kappa}}{2} = -2N_c \int -\frac{9|M_t|^2\Pi^0_{t_R}}{2p^2\Pi^0_{q_L}(\Pi^0_{t_R}+4\Pi^1_{t_R})^2}-9\left(\frac{\Pi^1_{t_R}}{\Pi^0_{t_R}+4\Pi^1_{t_R}}\right)^2,\\\nonumber
\frac{\lambda_h}{4} &= 2N_c \int \frac{9|M_t|^2\Pi^0_{t_R}}{2p^2\Pi^0_{q_L}(\Pi^0_{t_R}+4\Pi^1_{t_R})}+\frac{9|M_t|^2\Pi^1_{q_L}\Pi^0_{t_R}}{p^2(\Pi^0_{q_L})^2(\Pi^0_{t_R}+4\Pi^1_{t_R})^2}+\frac{36|M_t|^2\Pi^1_{q_L}\Pi^1_{t_R}}{p^2(\Pi^0_{q_L})^2(\Pi^0_{t_R}+4\Pi^1_{t_R})^2},\\\nonumber
\end{align}
where we have only retained the leading contributions in the expansion. All these integrals are understood over four-dimensional euclidean momentum. As we will discuss later on, in the natural vacuum of this potential the $h$ field is the only one taking a nonzero vev $v=\sqrt{\mu_h^2/\lambda_h}\simeq246$ GeV. In that case, the masses of these particles are given by
\begin{equation}
m_h^2 \simeq 2\lambda_h v^2, \qquad m_\eta^2 \simeq \mu_\eta^2+\lambda_{h\eta}v^2, \qquad m_{\kappa^\pm}^2 \simeq \mu_\kappa^2+\lambda_{h\kappa}v^2.
\end{equation}
In light of these equations and equations (\ref{formfactors}), $\kappa^\pm$ and $\eta$ become degenerate in mass. This degeneracy is broken by $\mathcal{O}(v/f)$ corrections when loops of $\bar{t}_L b_R$ are taken into account.

\section{Scalar couplings to SM particles}\label{couplings}
%
%
\begin{table}[t]
\begin{center}
\begin{tabular}{c|ccccc}\hline
Vertex & \textbf{Parameter} & \textbf{SILH} & \textbf{MCHM4} & \textbf{MCHM5} & \textbf{S7G2M}\\\hline
\raisebox{-.5\height}{\includegraphics[width=0.15\columnwidth]{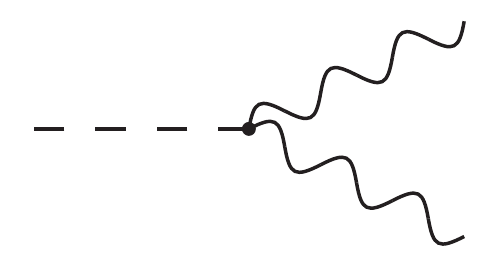}} &  $\mathbf{a}$ & $1-c_H\xi/2$ & $\sqrt{1-\xi}$ & $\sqrt{1-\xi}$ & $1-\frac{3}{8}\xi$\\\\
\raisebox{-.5\height}{\includegraphics[width=0.15\columnwidth]{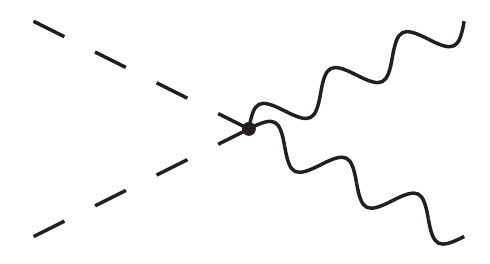}} & $\mathbf{b}$ & $1-2c_H\xi$ & $1-2\xi$ & $1-2\xi$ & $1-\frac{3}{2}\xi$\\\\
\raisebox{-.5\height}{\includegraphics[width=0.15\columnwidth]{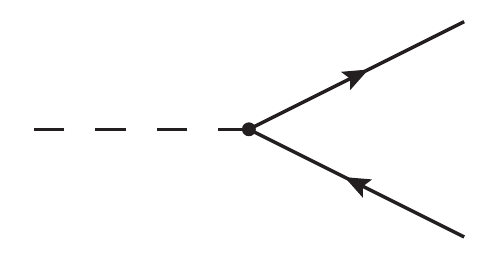}} & $\mathbf{c}$ & $1-(c_H/2+c_y)\xi$ & $\sqrt{1-\xi}$ & $\frac{1-2\xi}{\sqrt{1-\xi}}$ & $1-\frac{9}{8}\xi$\\\\
\raisebox{-.5\height}{\includegraphics[width=0.15\columnwidth]{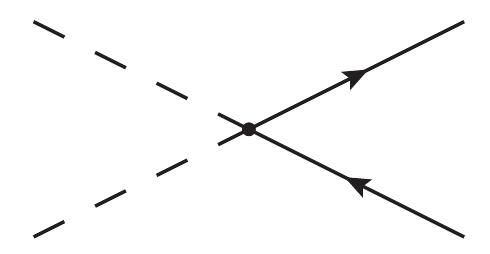}} & $\mathbf{c_2}$ & $-(c_H+3c_y)\xi/2$ & $-\xi/2$ & $-2\xi$ & $-\frac{3}{2}\xi$\\\\
\raisebox{-.5\height}{\includegraphics[width=0.15\columnwidth]{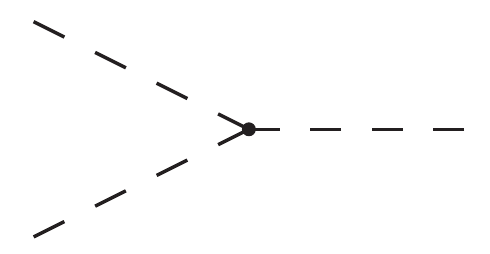}} & $\mathbf{d_3}$ & $1+(c_6-3c_H/2)\xi$ & $\sqrt{1-\xi}$ & $\frac{1-2\xi}{\sqrt{1-\xi}}$ & $1-\frac{9}{8}\xi$ \\\\\hline
\end{tabular}
\end{center}
\caption{Relevant Feynman diagrams and couplings for different structures of the CHM. The main part of the table has been extracted from Ref. \cite{Espinosa:2012qj}.}
\end{table}\label{comparison}
After gauging the SM subgroup of $G2\times U(1)_X$, the usual SM interactions appear, including the $hW^+W^-$ and $hZZ$ vertices, which were not present in the lagrangian of the last section. However, this field $h$ is not written in the canonical way, since after EWSB we get new contributions to the Higgs field kinetic term coming from the $\mathcal{O}_H$ operator $\mathcal{O}_H = \frac{c_H}{2f^2}(\partial_\mu(H^\dagger H))^2$ typical of the SILH model\cite{Giudice:2007fh}, where the $c_H$ parameter turns out to be $c_H = 3/4$ in this case. So, we should perform the following $h$ field redefinition in order to get it canonically normalized:
\begin{equation}\label{redef}
h = \left(1-\frac{3}{8}\xi+\mathcal{O}(\xi^2)\right)h_\text{phys}+v, \qquad \xi\equiv \frac{v^2}{f^2},
\end{equation}
where $f$ is the scale of new physics and $v\simeq 246$ GeV the electroweak scale. Thus, the coupling of $h$ to the gauge bosons $W$ and $Z$ change. After inserting $h_{\text{phys}}$ of equation (\ref{redef}) into the Yukawa lagrangian of equation (\ref{yukawa_end}), we get the coupling of the physical Higgs to the fermions. Similarly we get the rest of the couplings. We can write the Higgs sector lagrangian in the usual model-independent way\cite{Contino:2010mh}
\begin{align}
\mathcal{L}_{\text{EWSB}} &= \frac{v^4}{4}\text{Tr}\left(D_\mu\Sigma_{\text{ew}}^\dagger D^\mu\Sigma_\text{ew}\right)\times \left(1+2a\frac{h}{v}+b\frac{h^2}{v^2}+b_3\frac{h^3}{v^3}+\cdots\right)\\\nonumber
&-\frac{v}{\sqrt{2}}\left(\bar{t}^i_L\bar{b}^i_L\right)\Sigma_{\text{ew}}\left(1+c\frac{h}{v}+c_2\frac{h^2}{v^2}+\cdots\right)\left(y_{ij}^u t_R^j,y_{ij}^d b_R^j\right)^\text{T}+h.c.,
\end{align} 
and
\begin{equation}
V(h,\eta=0,\kappa=0) = \frac{1}{2}m_h^2h^2+d_3\left(\frac{m_h^2}{2v}\right)h^3+d_4\left(\frac{m_h^2}{8v^2}\right)h^4+\cdots
\end{equation}
Here $\Sigma_{\text{ew}}$ is different from $\Sigma$ of equation (\ref{sigma}), and parametrizes the coset manifold $SU(2)_L\times SU(2)_R/ SU(2)_V$ of the EWSB pattern. The corresponding parameters can be found in Table I for our model. For comparison we have also included the values in the MCHM4 and MCHM5 models. Note that we can always normalize the coset generators in a different way, which redefines the $f$ scale\cite{Giudice:2007fh}. So, $c_H$ can always take the same value in every CHM, while the ratio of $c_y/c_H$ is an actual prediction ($c_y$ is the coefficient in the operator $\frac{c_y y_f^{ij}}{f^2}H^\dagger H\bar{f}^i_L H f_R^j$). In our case, this ratio is the same as the MHCM5 one at this order in the $1/f$ expansion. Note also that in our construction, the leading contribution to $h\rightarrow gg$ and $h\rightarrow \gamma \gamma$ is fixed by group theory factors, independently of the composite spectrum\footnote{Naturalness arguments together with the recent light Higgs discovery, tend to prefer lighter fermion resonances for the third generation\cite{Redi:2012ha,Matsedonskyi:2012ym,Marzocca:2012zn,Pomarol:2012qf}.}. The reason in that the main contribution of the composite sector comes from the top-custodians resonances. If we promote the $T_R$ and $Q_L$ fields of equation (\ref{order2}) to complete $\mathbf{8}$ representations of $SO(7)$, we note that we can only construct the $SO(7)$-invariant $(\Sigma^T T_R)(\bar{Q}_L\Sigma)$. Since the top mass is sensibly larger than the Higgs mass, according to Ref. \cite{Azatov:2011qy} corrections to the $Hgg$ coupling will have no dependence on the masses of the composite partners, but only on some functions of $v$ as described in Table I.

Let us now briefly discuss the interactions of $\kappa^\pm$ and $\eta$ with the SM particles. Since the $\kappa^\pm$ are charged under $T^3_R$, they interact not only with the SM fermions through the lagrangian of equation (\ref{yukawa_end}), but also with the $Z$ and $\gamma$ bosons. These interactions are fixed by the gauge symmetry. They are given by the coupling of $A_\mu$ and $Z_\mu$ to the neutral current $J_\mu = i(\kappa^-\partial_\mu\kappa^+-\kappa^+\partial_\mu\kappa^-)$. The explicit trilinear interactions of $\kappa^\pm$ and $\eta$ scalars with the SM particles are shown in Table II. It is worth noting that the trilinear coupling of $\eta$ to the fermions disappears once $\theta$ is set to zero, when we recover the symmetry $\eta\rightarrow -\eta$.
\begin{table}[t]
\begin{center}
\begin{tabular}{l|c}\hline
 Feynman diagram \hspace{0.5cm} &  \textbf{Interaction term} \\\hline
\raisebox{-.5\height}{\includegraphics[width=0.2\columnwidth]{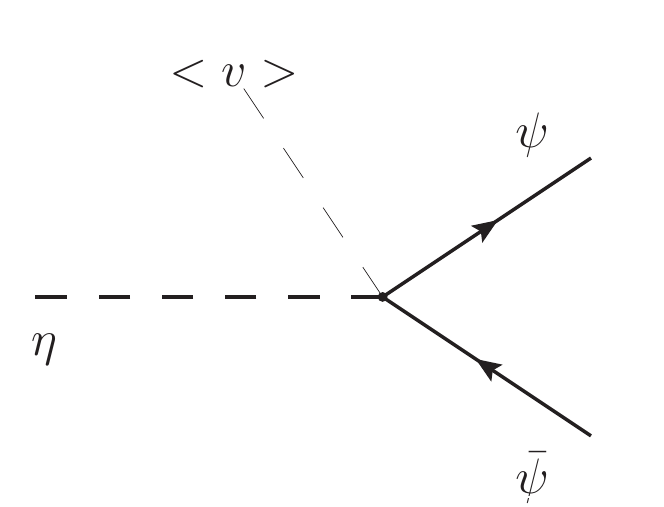}} & $\frac{3\tan{\theta}}{2\sqrt{3}} \frac{m_\psi}{f} \bar{\psi} (1+\gamma^5) \psi \eta$\\
\raisebox{-.5\height}{\includegraphics[width=0.2\columnwidth]{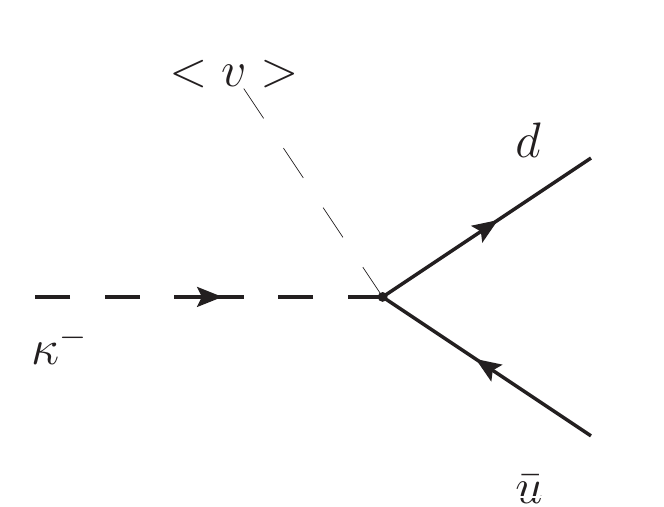}} & $\propto\epsilon\frac{v}{f}\bar{u}(1+\gamma^5)d \kappa^+$\\
\raisebox{-.5\height}{\includegraphics[width=0.2\columnwidth]{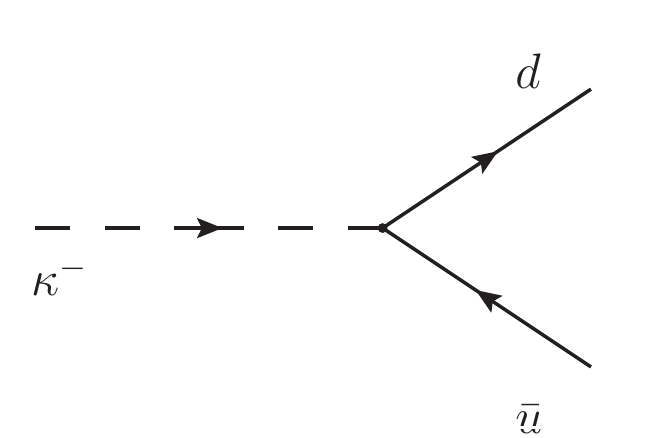}} & $\propto\frac{\epsilon}{f}\cos{\theta}\bar{u}(1-\gamma^5)\gamma^\mu\partial_\mu d\,\kappa^+$\\
\raisebox{-.5\height}{\includegraphics[width=0.2\columnwidth]{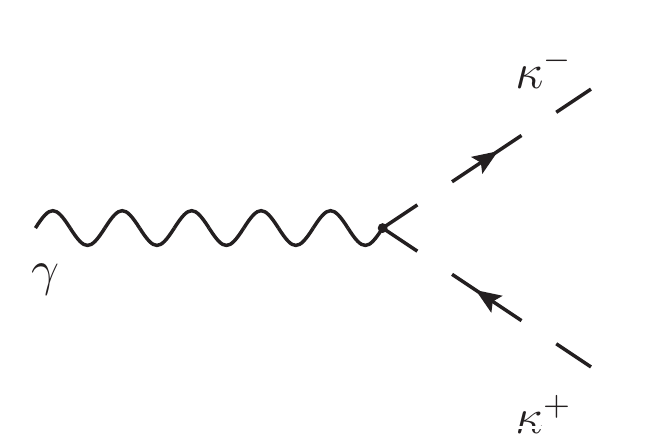}} & $i e\left(\kappa^-\partial_\mu\kappa^+-\kappa^+\partial_\mu\kappa^-\right)A^\mu $\\
\raisebox{-.5\height}{\includegraphics[width=0.2\columnwidth]{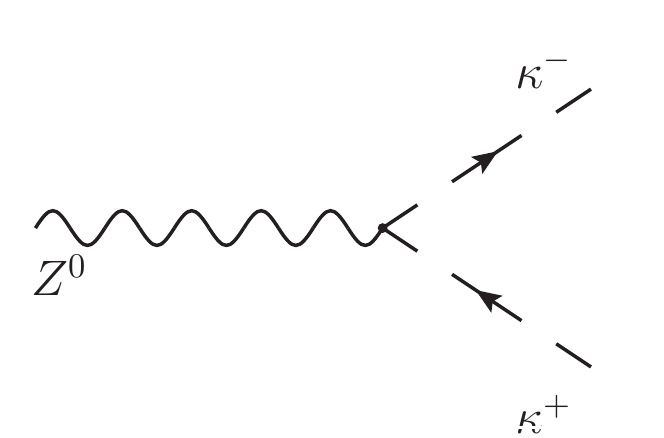}} & \hspace{0.2cm} $-i\frac{g s_w^2}{c_w}\left(\kappa^-\partial_\mu\kappa^+-\kappa^+\partial_\mu\kappa^-\right) Z^\mu$\\\\\hline
\end{tabular}
\caption{Trilinear interactions of $\kappa^\pm$ and $\eta$ with the SM particles. The interactions of $\kappa^\pm$ with the photon and the $Z$ are completely fixed by the gauge symmetry.}
\end{center}
\end{table}\label{otherinteractions}

\section{Phenomenological implications}\label{phenomenology}

Let us discuss now the main phenomenological implications of the extended composite scalar sector, including new contributions to $\Gamma(h\rightarrow\gamma\gamma)$, to DM and their LHC signals.

\subsection{The $h\rightarrow\gamma\gamma$ deviation}

Recent fits\cite{Montull:2012ik,Azatov:2012bz,Espinosa:2012ir,Corbett:2012dm,Carmi:2012yp} to a combination of ATLAS, CMS and Tevatron data on Higgs searches\cite{TEVNPH:2012ab,ATLAS-CONF-2012-019,CMS-PAS-HIG-12-008}, seem to point to an excess in $h\rightarrow\gamma\gamma$ events by a factor of $\sim1-2$ with respect to the SM prediction, while they agree pretty well in the rest of channels. Although the excess is not yet significant, it could mean an intriguing source of new physics. For this reason, many groups have proposed explanations to this discrepancy\cite{Djouadi:1998az,Petriello:2002uu,Han:2003gf,Chen:2006cs,Dermisek:2007fi,Low:2009nj,Low:2009di,Cacciapaglia:2009ky,Casagrande:2010si,Cheung:2011nv,Carena:2011aa,Cao:2011pg,Batell:2011pz,Arvanitaki:2011ck,Barger:2012hv,ArkaniHamed:2012kq,Arhrib:2012yv,Alves:2011kc,Lee:2012wz,Kearney:2012zi,Kanemura:2012rs,oglekar:2012vc,Dorsner:2012pp,Almeida:2012bq,Draper:2012xt,Akeroyd:2012ms,Dawson:2012di,Christensen:2012ei,Carena:2012xa,Delgado:2012sm,Chun:2012jw}, through the introduction of new uncolored (in order to not contribute to the gluon gluon fusion) particles which, when running in the loops, can increase the $\Gamma(h\rightarrow\gamma\gamma)$ width, while keeping the rest of the channels invariant. In this way, the branching ratio $\text{BR}(h\rightarrow\gamma\gamma)$ can be easily increased up to the current measurement $\sim 1.5\times\text{BR}^{SM}$ without conflicting with the others channels. These new particles could be both uncolored scalars or leptons. In any case, the new physics contribution would be suppressed by the mass scale of the new particle running in the loop. Thus, the lighter these particles are, the larger the contribution will be. This requirement can be successfully achieved in the context of CHM, where the Higgs boson, as well as possible new scalar states, arise as a pNGB of a new strongly interacting sector, being therefore naturally light. However, we should consider non minimal CHMs in order to get charged scalar states, like  $SO(6)/SO(4)\times SO(2)$ 2HDM. Spherical CHMs, based on $SO(n+1)/SO(n)$ could also explain this excess, whenever $n+1$ is such that possible anomalous representations appear. In that case, the Higgs boson could mix with the new extra singlet scalars which, although they are not charged, could still couple to pair of photons through $\eta F_{\mu\nu}F^{\mu\nu}$ interactions coming from anomalies in the strong sector\cite{Gripaios:2009pe,Montull:2012ik}. In the current model under study, however, a light charged scalar $\kappa^\pm$ arises naturally in the spectrum. As discussed before, its interactions with the Higgs boson $h$ and the singlet $\eta$ can be extracted from the generic potential in equation (\ref{potencial}),
\begin{figure}[t]
{\includegraphics[width=0.49\columnwidth]{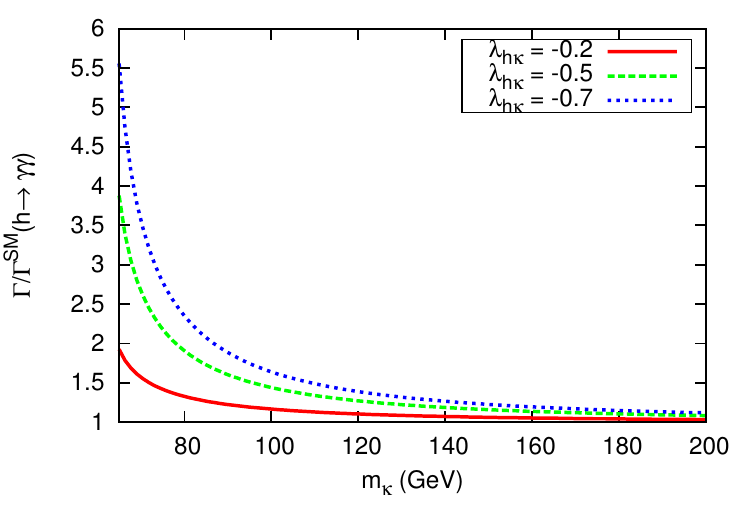}
\includegraphics[width=0.49\columnwidth]{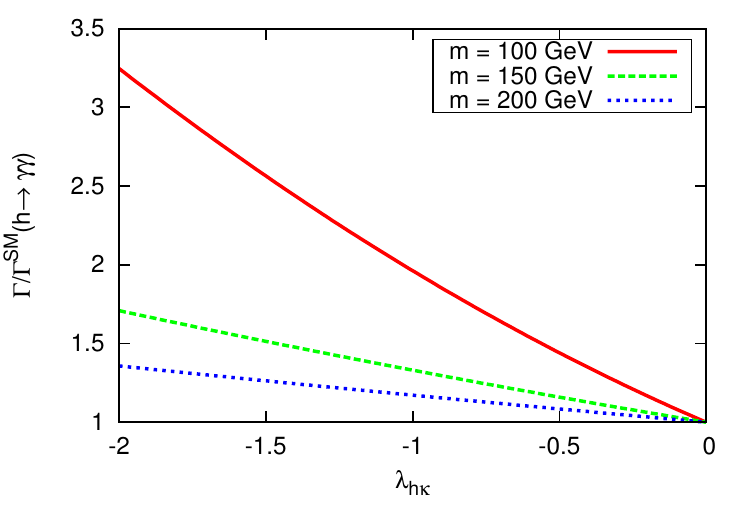}}
\caption{Left) $\Gamma/\Gamma^{SM}(h\rightarrow\gamma\gamma)$ in this model as a function of the mass of the new scalar $\kappa^\pm$ for different fixed $\lambda_{h\kappa}$ couplings. Right) $\Gamma/\Gamma^{SM}(h\rightarrow\gamma\gamma)$ as a function of $\lambda_{h\kappa}$ for different masses of $\kappa^\pm$. We are neglecting $\xi$ corrections in any case, so that the couplings are SM-like.}
\end{figure}\label{figgamma}
%
%
%
where all the mass parameters can be taken positive. In that case, the point $\langle h\rangle = \sqrt{\mu_h^2/\lambda_h}, \langle\eta\rangle=\langle\kappa\rangle = 0$ is necessarily a local minimum whenever the conditions $-\lambda_{h\eta}\mu_{h}^2 < \lambda_{h}\mu^2_\eta$ and $-\lambda_{h\kappa}\mu_{h}^2 < \lambda_{h}\mu^2_\kappa$ are satisfied. In particular, we see how this last condition can be successfully fulfilled with a negative $\lambda_{h\kappa}$ coupling. And this is, in fact, what we need to increase the $\gamma\gamma$ rate. After EWSB, the potential will give rise to the trilinear coupling $2\lambda_{h\kappa}vh\kappa^+\kappa^-$. In general, the addition of a uncolored singly charged scalar particle $S$ would modifiy the $\gamma\gamma$ width in the following way\cite{Carena:2012xa}:
\begin{equation}
 \Gamma(h\rightarrow\gamma\gamma) = \frac{\alpha^2 m_h^2}{1024\pi^3}\left|\frac{g_{hWW}}{m_W^2}A_1(\tau_W)+\frac{8g_{ht\bar{t}}}{3m_t}A_{1/2}(\tau_t)+\frac{g_{hSS}}{m_S^2}A_0(\tau_S)\right|^2,
\end{equation}
when the only relevant SM considered contributions come from $t$ and $W$ loops, and where $\alpha$, $m_h$, $m_W$ and $m_t$ are the fine-structure constant and the Higgs, $W$ and top mass, respectively. $g_{hWW}$ and $g_{htt}$ are the couplings constants of both $W$ and $t$ to the Higgs boson, which in the SM case become $g^2/2$ and $\lambda_t/\sqrt{2}$ respectively, while in CHM they receive deviations of order $\xi=v^2/f^2$ (see Table I). $m_S$ stands for the mass of the new scalar, $\tau_i \equiv 4m_i^2/m_h^2$ and $A_1$, $A_0$ and $A_{1/2}$ are defined in Appendix \ref{loop}.

In Figure 2, we show, for $\xi =0$ (SM-like couplings), the $\Gamma/\Gamma^{SM}(h\rightarrow\gamma\gamma)$ as a function of both $m_\kappa = \mu_\kappa^2+v^2\lambda_{h\kappa}$ and $\lambda_{h\kappa}$. $\Gamma(h\rightarrow\gamma\gamma)$ is reduced by 10\% for $\xi\sim 0.25$, but then the other channels become also modified\cite{Giudice:2007fh}.

\subsection{A Dark Matter Candidate}
\begin{figure}[t]\label{kprod}
\includegraphics[width=0.9\columnwidth]{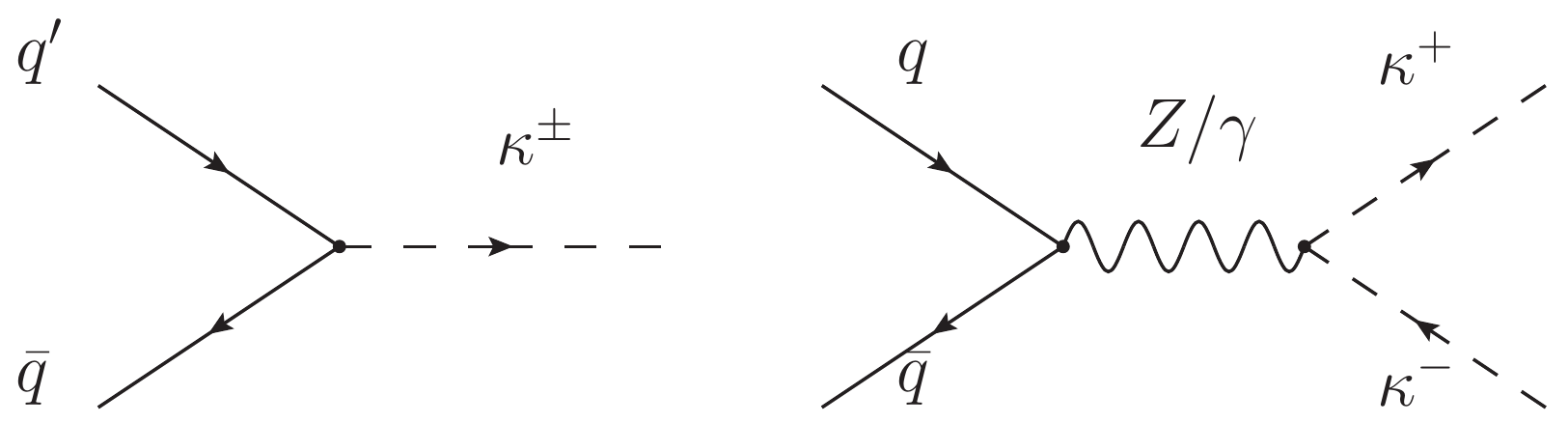}
\caption{Main $\kappa^\pm$ production channels at the LHC. The coupling in the left plot is of order $v/f$ and model-dependent, whereas the one in the diagram on the right panel is fixed by the gauge invariance.}
\end{figure}
As we see from the sigma model lagrangian of equation (\ref{sigmal}), there is a parity symmetry for $\eta\rightarrow -\eta$ that allows $\eta$ to be a dark matter candidate. If we were embedding the right-handed fermions in the remain neutral singlet of $SU(2)_L$, this symmetry would be broken explicitly, as mentioned in Section \ref{model}. Also interesting, contrary to the case of $SO(6)$ based CHMs (as $SO(6)/SO(5)$ of references \cite{Gripaios:2009pe,Frigerio:2012uc}), this model is free of anomalies, since $SO(7)$ is\cite{citeulike:712984}. Thus, we can not expect a WZW term breaking the parity symmetry\cite{AlvarezGaumé1985439}, and then the simplest CHM with natural dark matter candidates\footnote{Natural in the sense that the parity symmetry is not broken explicitly by the elementary sector and can not be broken through quantum anomalies.} would be $SO(7)/SO(6)$ and the current model $SO(7)/G2$. Phenomenologically speaking, the main differences would be the presence of two uncharged singlets in the first case, and then two possible DM candidates; and the presence of only one DM candidate and one charged singlet in the second. The DM implications of these models are very similar to that of the minimal realization studied in references \cite{YannMambrini2011,Yaguna:2008hd,Silveira1985136,McDonald:1993ex,Guo:2010hq,Farina2010329,Djouadi:2011aa,Burgess:2000yq}, with a nice exception (apart from the obvious problems associated to elementary light scalars). For large enough masses of the DM candidates, the CHM realizations become fully predictive, since the relic abundance turns out to be completely determined by derivative interactions, which are fixed by the coset structure\cite{Frigerio:2012uc}. In this particular model, of course, the interactions of $\kappa^\pm$ with $\eta$ could involve important differences with respect to Spherical CHMs. A detailed analysis of the different possibilities is beyond the scope of the present article and will be deferred to a future publication\cite{cerezo:2012di}.

\subsection{LHC Phenomenology}\label{lhcphenomenology}

The new zero charge singlet scalar, being a DM candidate, will be difficult to detect at the LHC. The case of $\kappa^\pm$, however, could be very different depending on its mass. For large masses, it could decay into $\bar{t}b$ pairs, and it could then be searched at the LHC in the $t\bar{t}b\bar{b}$ final state coming from $\kappa^+\kappa^-$ production, since its production cross section is completely determined, up to $\xi=v^2/f^2$ factors, by gauge interactions mediated by $Z/\gamma$. These large masses, however, would require a much larger coupling $\lambda_{h\kappa}$ in the operator $h^2 |\kappa^\pm|^2$ of equation (\ref{potencial}) to explain the observed discrepancy in $\text{BR}(h\rightarrow\gamma\gamma)$ ---as we have seen in Section \ref{couplings}, a negative sizable $\lambda_{h\kappa}$ coupling could increase the branching ratio up to $1.5\times\text{SM}$---. Moreover, searches of $W'$ decaying to a top and a bottom quarks can impose important constraints on the mass of the $\kappa^\pm$ boson. The most important experimental constraints on these searches come from CMS\cite{:2012sc}, and also from CDF\cite{PhysRevLett.90.081802} and D0\cite{Abazov:2008vj} experiments. CMS puts bounds on the mass of the $W'$ near the 2 TeV for a $W'tb$ coupling $g'\sim g_w$. In our case, the amplitude for the production of $\kappa^\pm$ and decay into $tb$ is proportional to $v^2/f^2 \sim 0.1$, to be compared with $g_w^2 \sim 0.5$, and could still be very significant (unless the corresponding couplings of Table II are small enough) depending on how the analyses affect the scalar signal compared to the vector one.

For lower masses, where the $\gamma\gamma$ excess can be successfully explained, $\kappa^\pm$ can no longer decay into on-shell $t$ and $b$ quarks, and the decay through virtual heavy quarks will be very small. Experiments at both LEP and LHC found limits on the mass $m_{H^\pm}$ of a charged SUSY-like scalar in the region $90$ GeV $<m_{H^\pm}<$ 160 GeV, depending on its properties\cite{Beringer:1900zz}. The decay into leptons and neutrinos is also strongly constrained by $W'$ searches\cite{Chatrchyan:2012meb,Chatrchyan:2012it,:2012hf,:2012dm}, and the decay into jets\cite{Chatrchyan:2011ns,Aad:2011fq} would be completely contaminated by the QCD background. So, very precise and dedicated analyses should be performed in other channels (resulting, for instance, from the decay of heavier resonances\cite{Barcelo:2011vk,Barcelo:2011wu,Bini:2011zb,Vignaroli:2012nf,Brooijmans:2012yi,Carmona:2012my}) to study the phenomenology of these scalars at the LHC. Some efforts in this direction can be found in Ref. \cite{Cagil:2012py, Huitu:2000ut,Akeroyd:2011zza}.

\section{Conclusions}\label{conclusions}

If the Higgs boson is a composite pNGB that arises from the breaking of $SO(7)$ to the subgroup $G2$, two extra light scalars ($SU(2)_L$ singlets, one charged and one neutral) are present in the spectrum. We have computed the most general effective lagrangian up to order $1/f^2$, with $f$ the decay constant in the strong sector, in the case that SM fermions mix with resonances transforming in the $\mathbf{8}$ spinorial representation of $SO(7)$. We have worked out the effective potential for the scalars induced by loops of the top quark, and discussed the pattern of EWSB. We have also discussed the interactions of the scalars with the SM fermions and gauge bosons and the phenomenology at the LHC. Loop corrections to $\Gamma(h\rightarrow\gamma\gamma)$ mediated by the charged scalar can make it significantly increase (up to $\sim 1.5$ times the SM width), explaining better the recent fits to the latests ATLAS, CMS and Tevatron results. In addition, the neutral scalar singlet could naturally be a DM candidate when the external elementary sources do not break the parity symmetry $\eta\rightarrow -\eta$. The new states can lead to a very interesting phenomenology that will be studied in a future publication.

\acknowledgments

It is a pleasure to thank F. del \'Aguila, R. Cerezo, J. M. Lizana, M. P\'erez-Victoria and J. Santiago for the very useful discussions. This work has been supported by the MICINN project FPA2010-17915, through the FPU programme and by Junta de Andaluc\'ia project FQM 101.


\appendix

\section{The Lie algebra of $G2$ and its embedding in $SO(7)$}\label{embeding}

The generators of $SO(7)$ in its $\mathbf{8}$ representation can be constructed out of the seven $\gamma$ matrices\cite{Evans:1994np}:
\begin{align}
 &\gamma_1 = i\sigma_2\otimes i\sigma_2\otimes i\sigma_2, \qquad \gamma_2 = \sigma_1\otimes i\sigma_2 \otimes 1,\\\nonumber
\qquad &\gamma_3 = i\sigma_2\otimes 1\otimes \sigma_1, \qquad \,\,\,\,\,\,\gamma_4 = -i\sigma_2\otimes 1\otimes \sigma_3,\\\nonumber
\qquad &\gamma_5 = 1\otimes \sigma_1\otimes i\sigma_2, \qquad \sigma_6 = -\sigma_3\otimes i\sigma_2\otimes 1, \qquad \gamma_7 = -1\otimes \sigma_3\otimes i\sigma_2.
\end{align}
where $\sigma_i$ represents the $i\text{th}$ Pauli matrix. Out of them, we can construct the generators as $J_{mn}=-J_{nm} = -[J_m,J_n]/4$. The $G2$ Lie algebra is then generated from the 14 generators $F_i$ and $M_i$ of the following set~\cite{citeulike:8040542}:
\begin{align}
\nonumber&F_1 = -\frac{i}{2}(J_{24}-J_{51}), \qquad M_1 = \frac{i}{\sqrt{3}}(J_{24}+J_{51}-2J_{73}), \qquad N_1 = \frac{i}{\sqrt{6}}(J_{24}+J_{51}+J_{73}),\\\nonumber
&F_2 = \frac{i}{2}(J_{54}-J_{12}), \qquad M_2 = -\frac{i}{\sqrt{12}}(J_{54}+J_{12}-2J_{67}), \qquad N_2 = \frac{i}{\sqrt{6}}(J_{54}+J_{12}+J_{67}),\\\nonumber
&F_3 = -\frac{i}{2}(J_{14}-J_{25}), \qquad M_3 = \frac{i}{\sqrt{12}}(J_{14}+J_{25}-2J_{36}), \qquad N_3 = \frac{i}{\sqrt{6}}(J_{14}+J_{25}+J_{36}),\\\nonumber
&F_4 = -\frac{i}{2}(J_{16}-J_{43}), \qquad M_4 = \frac{i}{\sqrt{12}}(J_{16}+J_{43}-2J_{72}), \qquad N_4 = \frac{i}{\sqrt{6}}(J_{16}+J_{43}+J_{72}),\\\nonumber
&F_5 = -\frac{i}{2}(J_{46}-J_{31}), \qquad M_5 = \frac{i}{\sqrt{12}}(J_{46}+J_{31}-2J_{57}), \qquad N_5 = \frac{i}{\sqrt{6}}(J_{46}+J_{31}+J_{57}),\\\nonumber
&F_6 = -\frac{i}{2}(J_{35}-J_{62}), \qquad M_6 = \frac{i}{\sqrt{12}}(J_{35}+J_{62}-2J_{71}), \qquad N_6 = \frac{i}{\sqrt{6}}(J_{35}+J_{62}+J_{71}),\\
&F_7 = \frac{i}{2}(J_{65}-J_{23}), \qquad M_7 = -\frac{i}{\sqrt{12}}(J_{65}+J_{23}-2J_{47}), \qquad N_7 = \frac{i}{\sqrt{6}}(J_{65}+J_{23}+J_{47}).
\end{align}
\section{Explicit expression of $\Pi$}\label{chorizo}

The explicit expression for $\Pi$ in the particular basis of Appendix \ref{embeding} is given by:

\begin{equation}
\Pi = \frac{i}{\sqrt{6}} \begin{pmatrix}0 & \frac{k_2}{2} & -\frac{k_1}{2} & \frac{k_2+\eta}{2} & -\frac{h}{2} & -\frac{\eta}{2} & 0 & -\frac{k_2+\eta}{2}\cr -\frac{k_2}{2} & 0 & -\frac{k_2-\eta}{2} & \frac{k_1}{2} & \frac{\eta}{2} & -\frac{2\,k_1+h}{2} & -\frac{k_2+\eta}{2} & -h\cr \frac{k_1}{2} & \frac{k_2-\eta}{2} & 0 & \frac{k_2}{2} & 0 & -\frac{k_2-\eta}{2} & -\frac{h}{2} & \frac{\eta}{2}\cr -\frac{k_2+\eta}{2} & -\frac{k_1}{2} & -\frac{k_2}{2} & 0 & -\frac{k_2-\eta}{2} & -h & -\frac{\eta}{2} & \frac{2\,k_1-h}{2}\cr \frac{h}{2} & -\frac{\eta}{2} & 0 & \frac{k_2-\eta}{2} & 0 & -\frac{k_2}{2} & \frac{k_1}{2} & -\frac{k_2-\eta}{2}\cr \frac{\eta}{2} & \frac{2\,k_1+h}{2} & \frac{k_2-\eta}{2} & h & \frac{k_2}{2} & 0 & \frac{k_2+\eta}{2} & -\frac{k_1}{2}\cr 0 & \frac{k_2+\eta}{2} & \frac{h}{2} & \frac{\eta}{2} & -\frac{k_1}{2} & -\frac{k_2+\eta}{2} & 0 & -\frac{k_2}{2}\cr \frac{k_2+\eta}{2} & h & -\frac{\eta}{2} & -\frac{2\,k_1-h}{2} & \frac{k_2-\eta}{2} & \frac{k_1}{2} & \frac{k_2}{2} & 0\end{pmatrix}.
\end{equation}

\vspace{0.5cm}

\section{Loop functions}\label{loop}

The loop functions $A_0$, $A_1$ and $A_{1/2}$ that appear in Section \ref{phenomenology} can be found in Appendix A of Ref. \cite{Carena:2012xa}:
\begin{align}
A_0(x)&=-x^2\left[x^{-1}-f(x^{-1})\right],\\\nonumber
A_1(x) & = -x^2\left[2x^{-2}+3x^{-1}+3(2x^{-1}-1)f(x^{-1})\right],\\\nonumber
A_{1/2}(x) &= 2x^2\left[x^{-1}+(x^{-1}-1)f(x^{-1})\right],
\end{align}
where
\begin{equation}
 f(x) = \arcsin^2{\sqrt{x}}
\end{equation}
whenever the Higgs mass is below the kinematic threshold of the loop particle: $m_{\text{loop}} \gtrsim 63$ GeV.

\bibliography{mybib}

\begin{thebibliography}{101}
\expandafter\ifx\csname natexlab\endcsname\relax\def\natexlab#1{#1}\fi
\expandafter\ifx\csname bibnamefont\endcsname\relax
  \def\bibnamefont#1{#1}\fi
\expandafter\ifx\csname bibfnamefont\endcsname\relax
  \def\bibfnamefont#1{#1}\fi
\expandafter\ifx\csname citenamefont\endcsname\relax
  \def\citenamefont#1{#1}\fi
\expandafter\ifx\csname url\endcsname\relax
  \def\url#1{\texttt{#1}}\fi
\expandafter\ifx\csname urlprefix\endcsname\relax\def\urlprefix{URL }\fi
\providecommand{\bibinfo}[2]{#2}
\providecommand{\eprint}[2][]{\url{#2}}

\bibitem[{\citenamefont{Kaplan and Georgi}(1984)}]{Kaplan:1983fs}
\bibinfo{author}{\bibfnamefont{D.~B.} \bibnamefont{Kaplan}} \bibnamefont{and}
  \bibinfo{author}{\bibfnamefont{H.}~\bibnamefont{Georgi}},
  \bibinfo{journal}{Phys.Lett.} \textbf{\bibinfo{volume}{B136}},
  \bibinfo{pages}{183} (\bibinfo{year}{1984}).

\bibitem[{\citenamefont{Kaplan et~al.}(1984)\citenamefont{Kaplan, Georgi, and
  Dimopoulos}}]{Kaplan:1983sm}
\bibinfo{author}{\bibfnamefont{D.~B.} \bibnamefont{Kaplan}},
  \bibinfo{author}{\bibfnamefont{H.}~\bibnamefont{Georgi}}, \bibnamefont{and}
  \bibinfo{author}{\bibfnamefont{S.}~\bibnamefont{Dimopoulos}},
  \bibinfo{journal}{Phys.Lett.} \textbf{\bibinfo{volume}{B136}},
  \bibinfo{pages}{187} (\bibinfo{year}{1984}).

\bibitem[{\citenamefont{Dimopoulos and Preskill}(1982)}]{Dimopoulos:1981xc}
\bibinfo{author}{\bibfnamefont{S.}~\bibnamefont{Dimopoulos}} \bibnamefont{and}
  \bibinfo{author}{\bibfnamefont{J.}~\bibnamefont{Preskill}},
  \bibinfo{journal}{Nucl.Phys.} \textbf{\bibinfo{volume}{B199}},
  \bibinfo{pages}{206} (\bibinfo{year}{1982}).

\bibitem[{\citenamefont{{TEVNPH (Tevatron New Phenomina and Higgs Working
  Group), CDF Collaboration and D0 Collaboration}}(2012)}]{TEVNPH:2012ab}
\bibinfo{author}{\bibnamefont{{TEVNPH (Tevatron New Phenomina and Higgs Working
  Group), CDF Collaboration and D0 Collaboration}}} (\bibinfo{year}{2012}),
  \eprint{1203.3774}.

\bibitem[{\citenamefont{{ATLAS Collaboration,
  ATLAS-CONF-2012-019}}(2012)}]{ATLAS-CONF-2012-019}
\bibinfo{author}{\bibnamefont{{ATLAS Collaboration, ATLAS-CONF-2012-019}}}
  (\bibinfo{year}{2012}).

\bibitem[{\citenamefont{{CMS Collaboration,
  CMS-PAS-HIG-12-008}}(2012)}]{CMS-PAS-HIG-12-008}
\bibinfo{author}{\bibnamefont{{CMS Collaboration, CMS-PAS-HIG-12-008}}}
  (\bibinfo{year}{2012}).

\bibitem[{\citenamefont{Agashe et~al.}(2005)\citenamefont{Agashe, Contino, and
  Pomarol}}]{Agashe:2004rs}
\bibinfo{author}{\bibfnamefont{K.}~\bibnamefont{Agashe}},
  \bibinfo{author}{\bibfnamefont{R.}~\bibnamefont{Contino}}, \bibnamefont{and}
  \bibinfo{author}{\bibfnamefont{A.}~\bibnamefont{Pomarol}},
  \bibinfo{journal}{Nucl.Phys.} \textbf{\bibinfo{volume}{B719}},
  \bibinfo{pages}{165} (\bibinfo{year}{2005}), \eprint{hep-ph/0412089}.

\bibitem[{\citenamefont{Contino et~al.}(2007)\citenamefont{Contino, Da~Rold,
  and Pomarol}}]{Contino:2006qr}
\bibinfo{author}{\bibfnamefont{R.}~\bibnamefont{Contino}},
  \bibinfo{author}{\bibfnamefont{L.}~\bibnamefont{Da~Rold}}, \bibnamefont{and}
  \bibinfo{author}{\bibfnamefont{A.}~\bibnamefont{Pomarol}},
  \bibinfo{journal}{Phys.Rev.} \textbf{\bibinfo{volume}{D75}},
  \bibinfo{pages}{055014} (\bibinfo{year}{2007}), \eprint{hep-ph/0612048}.

\bibitem[{\citenamefont{Gripaios et~al.}(2009)\citenamefont{Gripaios, Pomarol,
  Riva, and Serra}}]{Gripaios:2009pe}
\bibinfo{author}{\bibfnamefont{B.}~\bibnamefont{Gripaios}},
  \bibinfo{author}{\bibfnamefont{A.}~\bibnamefont{Pomarol}},
  \bibinfo{author}{\bibfnamefont{F.}~\bibnamefont{Riva}}, \bibnamefont{and}
  \bibinfo{author}{\bibfnamefont{J.}~\bibnamefont{Serra}},
  \bibinfo{journal}{JHEP} \textbf{\bibinfo{volume}{0904}}, \bibinfo{pages}{070}
  (\bibinfo{year}{2009}), \eprint{0902.1483}.

\bibitem[{\citenamefont{Mrazek et~al.}(2011)\citenamefont{Mrazek, Pomarol,
  Rattazzi, Redi, Serra et~al.}}]{Mrazek:2011iu}
\bibinfo{author}{\bibfnamefont{J.}~\bibnamefont{Mrazek}},
  \bibinfo{author}{\bibfnamefont{A.}~\bibnamefont{Pomarol}},
  \bibinfo{author}{\bibfnamefont{R.}~\bibnamefont{Rattazzi}},
  \bibinfo{author}{\bibfnamefont{M.}~\bibnamefont{Redi}},
  \bibinfo{author}{\bibfnamefont{J.}~\bibnamefont{Serra}},
  \bibnamefont{et~al.}, \bibinfo{journal}{Nucl.Phys.}
  \textbf{\bibinfo{volume}{B853}}, \bibinfo{pages}{1} (\bibinfo{year}{2011}),
  \eprint{1105.5403}.

\bibitem[{\citenamefont{Redi and Tesi}(2012)}]{Redi:2012ha}
\bibinfo{author}{\bibfnamefont{M.}~\bibnamefont{Redi}} \bibnamefont{and}
  \bibinfo{author}{\bibfnamefont{A.}~\bibnamefont{Tesi}}
  (\bibinfo{year}{2012}), \eprint{1205.0232}.

\bibitem[{\citenamefont{Bertuzzo et~al.}(2012)\citenamefont{Bertuzzo, Ray,
  de~Sandes, and Savoy}}]{Bertuzzo:2012ya}
\bibinfo{author}{\bibfnamefont{E.}~\bibnamefont{Bertuzzo}},
  \bibinfo{author}{\bibfnamefont{T.~S.} \bibnamefont{Ray}},
  \bibinfo{author}{\bibfnamefont{H.}~\bibnamefont{de~Sandes}},
  \bibnamefont{and} \bibinfo{author}{\bibfnamefont{C.~A.} \bibnamefont{Savoy}}
  (\bibinfo{year}{2012}), \eprint{1206.2623}.

\bibitem[{\citenamefont{Frigerio et~al.}(2012)\citenamefont{Frigerio, Pomarol,
  Riva, and Urbano}}]{Frigerio:2012uc}
\bibinfo{author}{\bibfnamefont{M.}~\bibnamefont{Frigerio}},
  \bibinfo{author}{\bibfnamefont{A.}~\bibnamefont{Pomarol}},
  \bibinfo{author}{\bibfnamefont{F.}~\bibnamefont{Riva}}, \bibnamefont{and}
  \bibinfo{author}{\bibfnamefont{A.}~\bibnamefont{Urbano}}
  (\bibinfo{year}{2012}), \eprint{1204.2808}.

\bibitem[{\citenamefont{Djouadi}(1998)}]{Djouadi:1998az}
\bibinfo{author}{\bibfnamefont{A.}~\bibnamefont{Djouadi}},
  \bibinfo{journal}{Phys.Lett.} \textbf{\bibinfo{volume}{B435}},
  \bibinfo{pages}{101} (\bibinfo{year}{1998}), \eprint{hep-ph/9806315}.

\bibitem[{\citenamefont{Petriello}(2002)}]{Petriello:2002uu}
\bibinfo{author}{\bibfnamefont{F.~J.} \bibnamefont{Petriello}},
  \bibinfo{journal}{JHEP} \textbf{\bibinfo{volume}{0205}}, \bibinfo{pages}{003}
  (\bibinfo{year}{2002}), \eprint{hep-ph/0204067}.

\bibitem[{\citenamefont{Han et~al.}(2003)\citenamefont{Han, Logan, McElrath,
  and Wang}}]{Han:2003gf}
\bibinfo{author}{\bibfnamefont{T.}~\bibnamefont{Han}},
  \bibinfo{author}{\bibfnamefont{H.~E.} \bibnamefont{Logan}},
  \bibinfo{author}{\bibfnamefont{B.}~\bibnamefont{McElrath}}, \bibnamefont{and}
  \bibinfo{author}{\bibfnamefont{L.-T.} \bibnamefont{Wang}},
  \bibinfo{journal}{Phys.Lett.} \textbf{\bibinfo{volume}{B563}},
  \bibinfo{pages}{191} (\bibinfo{year}{2003}), \eprint{hep-ph/0302188}.

\bibitem[{\citenamefont{Chen et~al.}(2006)\citenamefont{Chen, Tobe, and
  Yuan}}]{Chen:2006cs}
\bibinfo{author}{\bibfnamefont{C.-R.} \bibnamefont{Chen}},
  \bibinfo{author}{\bibfnamefont{K.}~\bibnamefont{Tobe}}, \bibnamefont{and}
  \bibinfo{author}{\bibfnamefont{C.~P.} \bibnamefont{Yuan}},
  \bibinfo{journal}{Phys.Lett.} \textbf{\bibinfo{volume}{B640}},
  \bibinfo{pages}{263} (\bibinfo{year}{2006}), \eprint{hep-ph/0602211}.

\bibitem[{\citenamefont{Dermisek and Low}(2008)}]{Dermisek:2007fi}
\bibinfo{author}{\bibfnamefont{R.}~\bibnamefont{Dermisek}} \bibnamefont{and}
  \bibinfo{author}{\bibfnamefont{I.}~\bibnamefont{Low}},
  \bibinfo{journal}{Phys.Rev.} \textbf{\bibinfo{volume}{D77}},
  \bibinfo{pages}{035012} (\bibinfo{year}{2008}), \eprint{hep-ph/0701235}.

\bibitem[{\citenamefont{Low and Shalgar}(2009)}]{Low:2009nj}
\bibinfo{author}{\bibfnamefont{I.}~\bibnamefont{Low}} \bibnamefont{and}
  \bibinfo{author}{\bibfnamefont{S.}~\bibnamefont{Shalgar}},
  \bibinfo{journal}{JHEP} \textbf{\bibinfo{volume}{0904}}, \bibinfo{pages}{091}
  (\bibinfo{year}{2009}), \eprint{0901.0266}.

\bibitem[{\citenamefont{Low et~al.}(2010)\citenamefont{Low, Rattazzi, and
  Vichi}}]{Low:2009di}
\bibinfo{author}{\bibfnamefont{I.}~\bibnamefont{Low}},
  \bibinfo{author}{\bibfnamefont{R.}~\bibnamefont{Rattazzi}}, \bibnamefont{and}
  \bibinfo{author}{\bibfnamefont{A.}~\bibnamefont{Vichi}},
  \bibinfo{journal}{JHEP} \textbf{\bibinfo{volume}{1004}}, \bibinfo{pages}{126}
  (\bibinfo{year}{2010}), \eprint{0907.5413}.

\bibitem[{\citenamefont{Cacciapaglia et~al.}(2009)\citenamefont{Cacciapaglia,
  Deandrea, and Llodra-Perez}}]{Cacciapaglia:2009ky}
\bibinfo{author}{\bibfnamefont{G.}~\bibnamefont{Cacciapaglia}},
  \bibinfo{author}{\bibfnamefont{A.}~\bibnamefont{Deandrea}}, \bibnamefont{and}
  \bibinfo{author}{\bibfnamefont{J.}~\bibnamefont{Llodra-Perez}},
  \bibinfo{journal}{JHEP} \textbf{\bibinfo{volume}{0906}}, \bibinfo{pages}{054}
  (\bibinfo{year}{2009}), \eprint{0901.0927}.

\bibitem[{\citenamefont{Casagrande et~al.}(2010)\citenamefont{Casagrande,
  Goertz, Haisch, Neubert, and Pfoh}}]{Casagrande:2010si}
\bibinfo{author}{\bibfnamefont{S.}~\bibnamefont{Casagrande}},
  \bibinfo{author}{\bibfnamefont{F.}~\bibnamefont{Goertz}},
  \bibinfo{author}{\bibfnamefont{U.}~\bibnamefont{Haisch}},
  \bibinfo{author}{\bibfnamefont{M.}~\bibnamefont{Neubert}}, \bibnamefont{and}
  \bibinfo{author}{\bibfnamefont{T.}~\bibnamefont{Pfoh}},
  \bibinfo{journal}{JHEP} \textbf{\bibinfo{volume}{1009}}, \bibinfo{pages}{014}
  (\bibinfo{year}{2010}), \eprint{1005.4315}.

\bibitem[{\citenamefont{Cheung and Yuan}(2012)}]{Cheung:2011nv}
\bibinfo{author}{\bibfnamefont{K.}~\bibnamefont{Cheung}} \bibnamefont{and}
  \bibinfo{author}{\bibfnamefont{T.-C.} \bibnamefont{Yuan}},
  \bibinfo{journal}{Phys.Rev.Lett.} \textbf{\bibinfo{volume}{108}},
  \bibinfo{pages}{141602} (\bibinfo{year}{2012}), \eprint{1112.4146}.

\bibitem[{\citenamefont{Carena et~al.}(2012{\natexlab{a}})\citenamefont{Carena,
  Gori, Shah, and Wagner}}]{Carena:2011aa}
\bibinfo{author}{\bibfnamefont{M.}~\bibnamefont{Carena}},
  \bibinfo{author}{\bibfnamefont{S.}~\bibnamefont{Gori}},
  \bibinfo{author}{\bibfnamefont{N.~R.} \bibnamefont{Shah}}, \bibnamefont{and}
  \bibinfo{author}{\bibfnamefont{C.~E.~M.} \bibnamefont{Wagner}},
  \bibinfo{journal}{JHEP} \textbf{\bibinfo{volume}{1203}}, \bibinfo{pages}{014}
  (\bibinfo{year}{2012}{\natexlab{a}}), \eprint{1112.3336}.

\bibitem[{\citenamefont{Cao et~al.}(2011)\citenamefont{Cao, Heng, Liu, and
  Yang}}]{Cao:2011pg}
\bibinfo{author}{\bibfnamefont{J.}~\bibnamefont{Cao}},
  \bibinfo{author}{\bibfnamefont{Z.}~\bibnamefont{Heng}},
  \bibinfo{author}{\bibfnamefont{T.}~\bibnamefont{Liu}}, \bibnamefont{and}
  \bibinfo{author}{\bibfnamefont{J.~M.} \bibnamefont{Yang}},
  \bibinfo{journal}{Phys.Lett.} \textbf{\bibinfo{volume}{B703}},
  \bibinfo{pages}{462} (\bibinfo{year}{2011}), \eprint{1103.0631}.

\bibitem[{\citenamefont{Batell et~al.}(2012)\citenamefont{Batell, Gori, and
  Wang}}]{Batell:2011pz}
\bibinfo{author}{\bibfnamefont{B.}~\bibnamefont{Batell}},
  \bibinfo{author}{\bibfnamefont{S.}~\bibnamefont{Gori}}, \bibnamefont{and}
  \bibinfo{author}{\bibfnamefont{L.-T.} \bibnamefont{Wang}},
  \bibinfo{journal}{JHEP} \textbf{\bibinfo{volume}{1206}}, \bibinfo{pages}{172}
  (\bibinfo{year}{2012}), \eprint{1112.5180}.

\bibitem[{\citenamefont{Arvanitaki and Villadoro}(2012)}]{Arvanitaki:2011ck}
\bibinfo{author}{\bibfnamefont{A.}~\bibnamefont{Arvanitaki}} \bibnamefont{and}
  \bibinfo{author}{\bibfnamefont{G.}~\bibnamefont{Villadoro}},
  \bibinfo{journal}{JHEP} \textbf{\bibinfo{volume}{1202}}, \bibinfo{pages}{144}
  (\bibinfo{year}{2012}), \eprint{1112.4835}.

\bibitem[{\citenamefont{Barger et~al.}(2012)\citenamefont{Barger, Ishida, and
  Keung}}]{Barger:2012hv}
\bibinfo{author}{\bibfnamefont{V.}~\bibnamefont{Barger}},
  \bibinfo{author}{\bibfnamefont{M.}~\bibnamefont{Ishida}}, \bibnamefont{and}
  \bibinfo{author}{\bibfnamefont{W.-Y.} \bibnamefont{Keung}},
  \bibinfo{journal}{Phys.Rev.Lett.} \textbf{\bibinfo{volume}{108}},
  \bibinfo{pages}{261801} (\bibinfo{year}{2012}), \eprint{1203.3456}.

\bibitem[{\citenamefont{Arkani-Hamed et~al.}(2012)\citenamefont{Arkani-Hamed,
  Blum, D{'}Agnolo, and Fan}}]{ArkaniHamed:2012kq}
\bibinfo{author}{\bibfnamefont{N.}~\bibnamefont{Arkani-Hamed}},
  \bibinfo{author}{\bibfnamefont{K.}~\bibnamefont{Blum}},
  \bibinfo{author}{\bibfnamefont{R.~T.} \bibnamefont{D{'}Agnolo}},
  \bibnamefont{and} \bibinfo{author}{\bibfnamefont{J.}~\bibnamefont{Fan}}
  (\bibinfo{year}{2012}), \eprint{1207.4482}.

\bibitem[{\citenamefont{Arhrib et~al.}(2012)\citenamefont{Arhrib, Benbrik, and
  Chen}}]{Arhrib:2012yv}
\bibinfo{author}{\bibfnamefont{A.}~\bibnamefont{Arhrib}},
  \bibinfo{author}{\bibfnamefont{R.}~\bibnamefont{Benbrik}}, \bibnamefont{and}
  \bibinfo{author}{\bibfnamefont{C.-H.} \bibnamefont{Chen}}
  (\bibinfo{year}{2012}), \eprint{1205.5536}.

\bibitem[{\citenamefont{Alves et~al.}(2011)\citenamefont{Alves,
  Ramirez~Barreto, Dias, de~S.~Pires, Queiroz et~al.}}]{Alves:2011kc}
\bibinfo{author}{\bibfnamefont{A.}~\bibnamefont{Alves}},
  \bibinfo{author}{\bibfnamefont{E.}~\bibnamefont{Ramirez~Barreto}},
  \bibinfo{author}{\bibfnamefont{A.~G.} \bibnamefont{Dias}},
  \bibinfo{author}{\bibfnamefont{C.~A.} \bibnamefont{de~S.~Pires}},
  \bibinfo{author}{\bibfnamefont{F.~S.} \bibnamefont{Queiroz}},
  \bibnamefont{et~al.}, \bibinfo{journal}{Phys.Rev.}
  \textbf{\bibinfo{volume}{D84}}, \bibinfo{pages}{115004}
  (\bibinfo{year}{2011}), \eprint{1109.0238}.

\bibitem[{\citenamefont{Lee et~al.}(2012)\citenamefont{Lee, Park, and
  Park}}]{Lee:2012wz}
\bibinfo{author}{\bibfnamefont{H.~M.} \bibnamefont{Lee}},
  \bibinfo{author}{\bibfnamefont{M.}~\bibnamefont{Park}}, \bibnamefont{and}
  \bibinfo{author}{\bibfnamefont{W.-I.} \bibnamefont{Park}}
  (\bibinfo{year}{2012}), \eprint{1209.1955}.

\bibitem[{\citenamefont{Kearney et~al.}(2012)\citenamefont{Kearney, Pierce, and
  Weiner}}]{Kearney:2012zi}
\bibinfo{author}{\bibfnamefont{J.}~\bibnamefont{Kearney}},
  \bibinfo{author}{\bibfnamefont{A.}~\bibnamefont{Pierce}}, \bibnamefont{and}
  \bibinfo{author}{\bibfnamefont{N.}~\bibnamefont{Weiner}}
  (\bibinfo{year}{2012}), \eprint{1207.7062}.

\bibitem[{\citenamefont{Kanemura and Yagyu}(2012)}]{Kanemura:2012rs}
\bibinfo{author}{\bibfnamefont{S.}~\bibnamefont{Kanemura}} \bibnamefont{and}
  \bibinfo{author}{\bibfnamefont{K.}~\bibnamefont{Yagyu}},
  \bibinfo{journal}{Phys.Rev.} \textbf{\bibinfo{volume}{D85}},
  \bibinfo{pages}{115009} (\bibinfo{year}{2012}), \eprint{1201.6287}.

\bibitem[{\citenamefont{Joglekar et~al.}(2012)\citenamefont{Joglekar,
  Schwaller, and Wagner}}]{oglekar:2012vc}
\bibinfo{author}{\bibfnamefont{A.}~\bibnamefont{Joglekar}},
  \bibinfo{author}{\bibfnamefont{P.}~\bibnamefont{Schwaller}},
  \bibnamefont{and} \bibinfo{author}{\bibfnamefont{C.~E.~M.}
  \bibnamefont{Wagner}} (\bibinfo{year}{2012}), \eprint{1207.4235}.

\bibitem[{\citenamefont{Dorsner et~al.}(2012)\citenamefont{Dorsner, Fajfer,
  Greljo, and Kamenik}}]{Dorsner:2012pp}
\bibinfo{author}{\bibfnamefont{I.}~\bibnamefont{Dorsner}},
  \bibinfo{author}{\bibfnamefont{S.}~\bibnamefont{Fajfer}},
  \bibinfo{author}{\bibfnamefont{A.}~\bibnamefont{Greljo}}, \bibnamefont{and}
  \bibinfo{author}{\bibfnamefont{J.~F.} \bibnamefont{Kamenik}}
  (\bibinfo{year}{2012}), \eprint{1208.1266}.

\bibitem[{\citenamefont{Almeida et~al.}(2012)\citenamefont{Almeida, Bertuzzo,
  Machado, and Funchal}}]{Almeida:2012bq}
\bibinfo{author}{\bibfnamefont{L.~G.} \bibnamefont{Almeida}},
  \bibinfo{author}{\bibfnamefont{E.}~\bibnamefont{Bertuzzo}},
  \bibinfo{author}{\bibfnamefont{P.~A.~N.} \bibnamefont{Machado}},
  \bibnamefont{and} \bibinfo{author}{\bibfnamefont{R.~Z.}
  \bibnamefont{Funchal}} (\bibinfo{year}{2012}), \eprint{1207.5254}.

\bibitem[{\citenamefont{Draper and McKeen}(2012)}]{Draper:2012xt}
\bibinfo{author}{\bibfnamefont{P.}~\bibnamefont{Draper}} \bibnamefont{and}
  \bibinfo{author}{\bibfnamefont{D.}~\bibnamefont{McKeen}},
  \bibinfo{journal}{Phys.Rev.} \textbf{\bibinfo{volume}{D85}},
  \bibinfo{pages}{115023} (\bibinfo{year}{2012}), \eprint{1204.1061}.

\bibitem[{\citenamefont{Akeroyd and Moretti}(2012)}]{Akeroyd:2012ms}
\bibinfo{author}{\bibfnamefont{A.~G.} \bibnamefont{Akeroyd}} \bibnamefont{and}
  \bibinfo{author}{\bibfnamefont{S.}~\bibnamefont{Moretti}},
  \bibinfo{journal}{Phys.Rev.} \textbf{\bibinfo{volume}{D86}},
  \bibinfo{pages}{035015} (\bibinfo{year}{2012}), \eprint{1206.0535}.

\bibitem[{\citenamefont{Dawson and Furlan}(2012)}]{Dawson:2012di}
\bibinfo{author}{\bibfnamefont{S.}~\bibnamefont{Dawson}} \bibnamefont{and}
  \bibinfo{author}{\bibfnamefont{E.}~\bibnamefont{Furlan}},
  \bibinfo{journal}{Phys.Rev.} \textbf{\bibinfo{volume}{D86}},
  \bibinfo{pages}{015021} (\bibinfo{year}{2012}), \eprint{1205.4733}.

\bibitem[{\citenamefont{Christensen et~al.}(2012)\citenamefont{Christensen,
  Han, and Su}}]{Christensen:2012ei}
\bibinfo{author}{\bibfnamefont{N.~D.} \bibnamefont{Christensen}},
  \bibinfo{author}{\bibfnamefont{T.}~\bibnamefont{Han}}, \bibnamefont{and}
  \bibinfo{author}{\bibfnamefont{S.}~\bibnamefont{Su}},
  \bibinfo{journal}{Phys.Rev.} \textbf{\bibinfo{volume}{D85}},
  \bibinfo{pages}{115018} (\bibinfo{year}{2012}), \eprint{1203.3207}.

\bibitem[{\citenamefont{Carena et~al.}(2012{\natexlab{b}})\citenamefont{Carena,
  Low, and Wagner}}]{Carena:2012xa}
\bibinfo{author}{\bibfnamefont{M.}~\bibnamefont{Carena}},
  \bibinfo{author}{\bibfnamefont{I.}~\bibnamefont{Low}}, \bibnamefont{and}
  \bibinfo{author}{\bibfnamefont{C.~E.~M.} \bibnamefont{Wagner}},
  \bibinfo{journal}{JHEP} \textbf{\bibinfo{volume}{1208}}, \bibinfo{pages}{060}
  (\bibinfo{year}{2012}{\natexlab{b}}), \eprint{1206.1082}.

\bibitem[{\citenamefont{Delgado et~al.}(2012)\citenamefont{Delgado, Nardini,
  and Quiros}}]{Delgado:2012sm}
\bibinfo{author}{\bibfnamefont{A.}~\bibnamefont{Delgado}},
  \bibinfo{author}{\bibfnamefont{G.}~\bibnamefont{Nardini}}, \bibnamefont{and}
  \bibinfo{author}{\bibfnamefont{M.}~\bibnamefont{Quiros}}
  (\bibinfo{year}{2012}), \eprint{1207.6596}.

\bibitem[{\citenamefont{Chun et~al.}(2012)\citenamefont{Chun, Lee, and
  Sharma}}]{Chun:2012jw}
\bibinfo{author}{\bibfnamefont{E.~J.} \bibnamefont{Chun}},
  \bibinfo{author}{\bibfnamefont{H.~M.} \bibnamefont{Lee}}, \bibnamefont{and}
  \bibinfo{author}{\bibfnamefont{P.}~\bibnamefont{Sharma}}
  (\bibinfo{year}{2012}), \eprint{1209.1303}.

\bibitem[{\citenamefont{Carmona et~al.}(2012)\citenamefont{Carmona, Chala, and
  Santiago}}]{Carmona:2012my}
\bibinfo{author}{\bibfnamefont{A.}~\bibnamefont{Carmona}},
  \bibinfo{author}{\bibfnamefont{M.}~\bibnamefont{Chala}}, \bibnamefont{and}
  \bibinfo{author}{\bibfnamefont{J.}~\bibnamefont{Santiago}}
  (\bibinfo{year}{2012}), \bibinfo{note}{27 pages, 22 figures},
  \eprint{1205.2378}.

\bibitem[{\citenamefont{Vignaroli}(2012)}]{Vignaroli:2012nf}
\bibinfo{author}{\bibfnamefont{N.}~\bibnamefont{Vignaroli}}
  (\bibinfo{year}{2012}), \eprint{1207.0830}.

\bibitem[{\citenamefont{Barcelo
  et~al.}(2012{\natexlab{a}})\citenamefont{Barcelo, Carmona, Chala, Masip, and
  Santiago}}]{Barcelo:2011wu}
\bibinfo{author}{\bibfnamefont{R.}~\bibnamefont{Barcelo}},
  \bibinfo{author}{\bibfnamefont{A.}~\bibnamefont{Carmona}},
  \bibinfo{author}{\bibfnamefont{M.}~\bibnamefont{Chala}},
  \bibinfo{author}{\bibfnamefont{M.}~\bibnamefont{Masip}}, \bibnamefont{and}
  \bibinfo{author}{\bibfnamefont{J.}~\bibnamefont{Santiago}},
  \bibinfo{journal}{Nucl.Phys.} \textbf{\bibinfo{volume}{B857}},
  \bibinfo{pages}{172} (\bibinfo{year}{2012}{\natexlab{a}}), \bibinfo{note}{17
  pages, 8 figures. Version 2: references added}, \eprint{1110.5914}.

\bibitem[{\citenamefont{Bini et~al.}(2012)\citenamefont{Bini, Contino, and
  Vignaroli}}]{Bini:2011zb}
\bibinfo{author}{\bibfnamefont{C.}~\bibnamefont{Bini}},
  \bibinfo{author}{\bibfnamefont{R.}~\bibnamefont{Contino}}, \bibnamefont{and}
  \bibinfo{author}{\bibfnamefont{N.}~\bibnamefont{Vignaroli}},
  \bibinfo{journal}{JHEP} \textbf{\bibinfo{volume}{1201}}, \bibinfo{pages}{157}
  (\bibinfo{year}{2012}), \eprint{1110.6058}.

\bibitem[{\citenamefont{Brooijmans et~al.}(2012)\citenamefont{Brooijmans,
  Gripaios, Moortgat, Santiago, Skands et~al.}}]{Brooijmans:2012yi}
\bibinfo{author}{\bibfnamefont{G.}~\bibnamefont{Brooijmans}},
  \bibinfo{author}{\bibfnamefont{B.}~\bibnamefont{Gripaios}},
  \bibinfo{author}{\bibfnamefont{F.}~\bibnamefont{Moortgat}},
  \bibinfo{author}{\bibfnamefont{J.}~\bibnamefont{Santiago}},
  \bibinfo{author}{\bibfnamefont{P.}~\bibnamefont{Skands}},
  \bibnamefont{et~al.} (\bibinfo{year}{2012}), \eprint{1203.1488}.

\bibitem[{\citenamefont{Contino et~al.}(2010)\citenamefont{Contino, Grojean,
  Moretti, Piccinini, and Rattazzi}}]{Contino:2010mh}
\bibinfo{author}{\bibfnamefont{R.}~\bibnamefont{Contino}},
  \bibinfo{author}{\bibfnamefont{C.}~\bibnamefont{Grojean}},
  \bibinfo{author}{\bibfnamefont{M.}~\bibnamefont{Moretti}},
  \bibinfo{author}{\bibfnamefont{F.}~\bibnamefont{Piccinini}},
  \bibnamefont{and} \bibinfo{author}{\bibfnamefont{R.}~\bibnamefont{Rattazzi}},
  \bibinfo{journal}{JHEP} \textbf{\bibinfo{volume}{1005}}, \bibinfo{pages}{089}
  (\bibinfo{year}{2010}), \eprint{1002.1011}.

\bibitem[{\citenamefont{Grober and Muhlleitner}(2011)}]{Grober:2010yv}
\bibinfo{author}{\bibfnamefont{R.}~\bibnamefont{Grober}} \bibnamefont{and}
  \bibinfo{author}{\bibfnamefont{M.}~\bibnamefont{Muhlleitner}},
  \bibinfo{journal}{JHEP} \textbf{\bibinfo{volume}{1106}}, \bibinfo{pages}{020}
  (\bibinfo{year}{2011}), \eprint{1012.1562}.

\bibitem[{\citenamefont{Contino et~al.}(2012)\citenamefont{Contino, Ghezzi,
  Moretti, Panico, Piccinini et~al.}}]{Contino:2012xk}
\bibinfo{author}{\bibfnamefont{R.}~\bibnamefont{Contino}},
  \bibinfo{author}{\bibfnamefont{M.}~\bibnamefont{Ghezzi}},
  \bibinfo{author}{\bibfnamefont{M.}~\bibnamefont{Moretti}},
  \bibinfo{author}{\bibfnamefont{G.}~\bibnamefont{Panico}},
  \bibinfo{author}{\bibfnamefont{F.}~\bibnamefont{Piccinini}},
  \bibnamefont{et~al.}, \bibinfo{journal}{JHEP}
  \textbf{\bibinfo{volume}{1208}}, \bibinfo{pages}{154} (\bibinfo{year}{2012}),
  \eprint{1205.5444}.

\bibitem[{\citenamefont{Gillioz et~al.}(2012)\citenamefont{Gillioz, Grober,
  Grojean, Muhlleitner, and Salvioni}}]{Gillioz:2012se}
\bibinfo{author}{\bibfnamefont{M.}~\bibnamefont{Gillioz}},
  \bibinfo{author}{\bibfnamefont{R.}~\bibnamefont{Grober}},
  \bibinfo{author}{\bibfnamefont{C.}~\bibnamefont{Grojean}},
  \bibinfo{author}{\bibfnamefont{M.}~\bibnamefont{Muhlleitner}},
  \bibnamefont{and} \bibinfo{author}{\bibfnamefont{E.}~\bibnamefont{Salvioni}}
  (\bibinfo{year}{2012}), \eprint{1206.7120}.

\bibitem[{\citenamefont{G{\"{u}}naydin and
  G{\"{u}}rsey}(1973)}]{citeulike:8040542}
\bibinfo{author}{\bibfnamefont{M.}~\bibnamefont{G{\"{u}}naydin}}
  \bibnamefont{and}
  \bibinfo{author}{\bibfnamefont{F.}~\bibnamefont{G{\"{u}}rsey}},
  \bibinfo{journal}{Journal of Mathematical Physics}
  \textbf{\bibinfo{volume}{14}}, \bibinfo{pages}{1651} (\bibinfo{year}{1973}),
  \urlprefix\url{http://dx.doi.org/10.1063/1.1666240}.

\bibitem[{\citenamefont{Evans}(1994)}]{Evans:1994np}
\bibinfo{author}{\bibfnamefont{J.~M.} \bibnamefont{Evans}},
  \bibinfo{journal}{Phys.Lett.} \textbf{\bibinfo{volume}{B334}},
  \bibinfo{pages}{105} (\bibinfo{year}{1994}), \eprint{hep-th/9404190}.

\bibitem[{\citenamefont{Gunaydin and Ketov}(1996)}]{Gunaydin:1995as}
\bibinfo{author}{\bibfnamefont{M.}~\bibnamefont{Gunaydin}} \bibnamefont{and}
  \bibinfo{author}{\bibfnamefont{S.~V.} \bibnamefont{Ketov}},
  \bibinfo{journal}{Nucl.Phys.} \textbf{\bibinfo{volume}{B467}},
  \bibinfo{pages}{215} (\bibinfo{year}{1996}), \eprint{hep-th/9601072}.

\bibitem[{\citenamefont{Agashe et~al.}(2006)\citenamefont{Agashe, Contino,
  Da~Rold, and Pomarol}}]{Agashe:2006at}
\bibinfo{author}{\bibfnamefont{K.}~\bibnamefont{Agashe}},
  \bibinfo{author}{\bibfnamefont{R.}~\bibnamefont{Contino}},
  \bibinfo{author}{\bibfnamefont{L.}~\bibnamefont{Da~Rold}}, \bibnamefont{and}
  \bibinfo{author}{\bibfnamefont{A.}~\bibnamefont{Pomarol}},
  \bibinfo{journal}{Phys.Lett.} \textbf{\bibinfo{volume}{B641}},
  \bibinfo{pages}{62} (\bibinfo{year}{2006}), \eprint{hep-ph/0605341}.

\bibitem[{\citenamefont{R{\'{u}}jula et~al.}(1990)\citenamefont{R{\'{u}}jula,
  Glashow, and Sarid}}]{DeRújula1990173}
\bibinfo{author}{\bibfnamefont{A.~D.} \bibnamefont{R{\'{u}}jula}},
  \bibinfo{author}{\bibfnamefont{S.~L.} \bibnamefont{Glashow}},
  \bibnamefont{and} \bibinfo{author}{\bibfnamefont{U.}~\bibnamefont{Sarid}},
  \bibinfo{journal}{Nuclear Physics B} \textbf{\bibinfo{volume}{333}},
  \bibinfo{pages}{173} (\bibinfo{year}{1990}), ISSN \bibinfo{issn}{0550-3213},
  \urlprefix\url{http://www.sciencedirect.com/science/article/pii/055032139090%
2275}.

\bibitem[{\citenamefont{Dimopoulos et~al.}(1990)\citenamefont{Dimopoulos,
  Eichler, Esmailzadeh, and Starkman}}]{PhysRevD.41.2388}
\bibinfo{author}{\bibfnamefont{S.}~\bibnamefont{Dimopoulos}},
  \bibinfo{author}{\bibfnamefont{D.}~\bibnamefont{Eichler}},
  \bibinfo{author}{\bibfnamefont{R.}~\bibnamefont{Esmailzadeh}},
  \bibnamefont{and} \bibinfo{author}{\bibfnamefont{G.~D.}
  \bibnamefont{Starkman}}, \bibinfo{journal}{Phys. Rev. D}
  \textbf{\bibinfo{volume}{41}}, \bibinfo{pages}{2388} (\bibinfo{year}{1990}),
  \urlprefix\url{http://link.aps.org/doi/10.1103/PhysRevD.41.2388}.

\bibitem[{\citenamefont{Chivukula et~al.}(1990)\citenamefont{Chivukula, Cohen,
  Dimopoulos, and Walker}}]{PhysRevLett.65.957}
\bibinfo{author}{\bibfnamefont{R.~S.} \bibnamefont{Chivukula}},
  \bibinfo{author}{\bibfnamefont{A.~G.} \bibnamefont{Cohen}},
  \bibinfo{author}{\bibfnamefont{S.}~\bibnamefont{Dimopoulos}},
  \bibnamefont{and} \bibinfo{author}{\bibfnamefont{T.~P.}
  \bibnamefont{Walker}}, \bibinfo{journal}{Phys. Rev. Lett.}
  \textbf{\bibinfo{volume}{65}}, \bibinfo{pages}{957} (\bibinfo{year}{1990}),
  \urlprefix\url{http://link.aps.org/doi/10.1103/PhysRevLett.65.957}.

\bibitem[{\citenamefont{Gould et~al.}(1990)\citenamefont{Gould, Draine, Romani,
  and Nussinov}}]{Gould:1989gw}
\bibinfo{author}{\bibfnamefont{A.}~\bibnamefont{Gould}},
  \bibinfo{author}{\bibfnamefont{B.~T.} \bibnamefont{Draine}},
  \bibinfo{author}{\bibfnamefont{R.~W.} \bibnamefont{Romani}},
  \bibnamefont{and} \bibinfo{author}{\bibfnamefont{S.}~\bibnamefont{Nussinov}},
  \bibinfo{journal}{Phys.Lett.} \textbf{\bibinfo{volume}{B238}},
  \bibinfo{pages}{337} (\bibinfo{year}{1990}).

\bibitem[{\citenamefont{Coleman and Weinberg}(1973)}]{PhysRevD.7.1888}
\bibinfo{author}{\bibfnamefont{S.}~\bibnamefont{Coleman}} \bibnamefont{and}
  \bibinfo{author}{\bibfnamefont{E.}~\bibnamefont{Weinberg}},
  \bibinfo{journal}{Phys. Rev. D} \textbf{\bibinfo{volume}{7}},
  \bibinfo{pages}{1888} (\bibinfo{year}{1973}),
  \urlprefix\url{http://link.aps.org/doi/10.1103/PhysRevD.7.1888}.

\bibitem[{\citenamefont{Witten}(1983)}]{PhysRevLett.51.2351}
\bibinfo{author}{\bibfnamefont{E.}~\bibnamefont{Witten}},
  \bibinfo{journal}{Phys. Rev. Lett.} \textbf{\bibinfo{volume}{51}},
  \bibinfo{pages}{2351} (\bibinfo{year}{1983}),
  \urlprefix\url{http://link.aps.org/doi/10.1103/PhysRevLett.51.2351}.

\bibitem[{\citenamefont{Jackiw}(1974)}]{PhysRevD.9.1686}
\bibinfo{author}{\bibfnamefont{R.}~\bibnamefont{Jackiw}},
  \bibinfo{journal}{Phys. Rev. D} \textbf{\bibinfo{volume}{9}},
  \bibinfo{pages}{1686} (\bibinfo{year}{1974}),
  \urlprefix\url{http://link.aps.org/doi/10.1103/PhysRevD.9.1686}.

\bibitem[{\citenamefont{Weinberg}(1967)}]{PhysRevLett.18.507}
\bibinfo{author}{\bibfnamefont{S.}~\bibnamefont{Weinberg}},
  \bibinfo{journal}{Phys. Rev. Lett.} \textbf{\bibinfo{volume}{18}},
  \bibinfo{pages}{507} (\bibinfo{year}{1967}),
  \urlprefix\url{http://link.aps.org/doi/10.1103/PhysRevLett.18.507}.

\bibitem[{\citenamefont{Marzocca et~al.}(2012)\citenamefont{Marzocca, Serone,
  and Shu}}]{Marzocca:2012zn}
\bibinfo{author}{\bibfnamefont{D.}~\bibnamefont{Marzocca}},
  \bibinfo{author}{\bibfnamefont{M.}~\bibnamefont{Serone}}, \bibnamefont{and}
  \bibinfo{author}{\bibfnamefont{J.}~\bibnamefont{Shu}} (\bibinfo{year}{2012}),
  \eprint{1205.0770}.

\bibitem[{\citenamefont{Pomarol and Riva}(2012)}]{Pomarol:2012qf}
\bibinfo{author}{\bibfnamefont{A.}~\bibnamefont{Pomarol}} \bibnamefont{and}
  \bibinfo{author}{\bibfnamefont{F.}~\bibnamefont{Riva}},
  \bibinfo{journal}{JHEP} \textbf{\bibinfo{volume}{1208}}, \bibinfo{pages}{135}
  (\bibinfo{year}{2012}), \eprint{1205.6434}.

\bibitem[{\citenamefont{Espinosa
  et~al.}(2012{\natexlab{a}})\citenamefont{Espinosa, Grojean, and
  Muhlleitner}}]{Espinosa:2012qj}
\bibinfo{author}{\bibfnamefont{J.~R.} \bibnamefont{Espinosa}},
  \bibinfo{author}{\bibfnamefont{C.}~\bibnamefont{Grojean}}, \bibnamefont{and}
  \bibinfo{author}{\bibfnamefont{M.}~\bibnamefont{Muhlleitner}}
  (\bibinfo{year}{2012}{\natexlab{a}}), \bibinfo{note}{6 pages. Contribution to
  the proceedings of Hadron Collider Physics Symposium 2011, Paris Nov. 14-18},
  \eprint{1202.1286}.

\bibitem[{\citenamefont{Giudice et~al.}(2007)\citenamefont{Giudice, Grojean,
  Pomarol, and Rattazzi}}]{Giudice:2007fh}
\bibinfo{author}{\bibfnamefont{G.~F.} \bibnamefont{Giudice}},
  \bibinfo{author}{\bibfnamefont{C.}~\bibnamefont{Grojean}},
  \bibinfo{author}{\bibfnamefont{A.}~\bibnamefont{Pomarol}}, \bibnamefont{and}
  \bibinfo{author}{\bibfnamefont{R.}~\bibnamefont{Rattazzi}},
  \bibinfo{journal}{JHEP} \textbf{\bibinfo{volume}{0706}}, \bibinfo{pages}{045}
  (\bibinfo{year}{2007}), \eprint{hep-ph/0703164}.

\bibitem[{\citenamefont{Matsedonskyi et~al.}(2012)\citenamefont{Matsedonskyi,
  Panico, and Wulzer}}]{Matsedonskyi:2012ym}
\bibinfo{author}{\bibfnamefont{O.}~\bibnamefont{Matsedonskyi}},
  \bibinfo{author}{\bibfnamefont{G.}~\bibnamefont{Panico}}, \bibnamefont{and}
  \bibinfo{author}{\bibfnamefont{A.}~\bibnamefont{Wulzer}}
  (\bibinfo{year}{2012}), \eprint{1204.6333}.

\bibitem[{\citenamefont{Azatov and Galloway}(2012)}]{Azatov:2011qy}
\bibinfo{author}{\bibfnamefont{A.}~\bibnamefont{Azatov}} \bibnamefont{and}
  \bibinfo{author}{\bibfnamefont{J.}~\bibnamefont{Galloway}},
  \bibinfo{journal}{Phys.Rev.} \textbf{\bibinfo{volume}{D85}},
  \bibinfo{pages}{055013} (\bibinfo{year}{2012}), \eprint{1110.5646}.

\bibitem[{\citenamefont{Montull and Riva}(2012)}]{Montull:2012ik}
\bibinfo{author}{\bibfnamefont{M.}~\bibnamefont{Montull}} \bibnamefont{and}
  \bibinfo{author}{\bibfnamefont{F.}~\bibnamefont{Riva}}
  (\bibinfo{year}{2012}), \eprint{1207.1716}.

\bibitem[{\citenamefont{Azatov et~al.}(2012)\citenamefont{Azatov, Contino, and
  Galloway}}]{Azatov:2012bz}
\bibinfo{author}{\bibfnamefont{A.}~\bibnamefont{Azatov}},
  \bibinfo{author}{\bibfnamefont{R.}~\bibnamefont{Contino}}, \bibnamefont{and}
  \bibinfo{author}{\bibfnamefont{J.}~\bibnamefont{Galloway}}
  (\bibinfo{year}{2012}), \eprint{1202.3415}.

\bibitem[{\citenamefont{Espinosa
  et~al.}(2012{\natexlab{b}})\citenamefont{Espinosa, Grojean, Muhlleitner, and
  Trott}}]{Espinosa:2012ir}
\bibinfo{author}{\bibfnamefont{J.~R.} \bibnamefont{Espinosa}},
  \bibinfo{author}{\bibfnamefont{C.}~\bibnamefont{Grojean}},
  \bibinfo{author}{\bibfnamefont{M.}~\bibnamefont{Muhlleitner}},
  \bibnamefont{and} \bibinfo{author}{\bibfnamefont{M.}~\bibnamefont{Trott}}
  (\bibinfo{year}{2012}{\natexlab{b}}), \eprint{1202.3697}.

\bibitem[{\citenamefont{Corbett et~al.}(2012)\citenamefont{Corbett, Eboli,
  Gonzalez-Fraile, and Gonzalez-Garcia}}]{Corbett:2012dm}
\bibinfo{author}{\bibfnamefont{T.}~\bibnamefont{Corbett}},
  \bibinfo{author}{\bibfnamefont{O.~J.~P.} \bibnamefont{Eboli}},
  \bibinfo{author}{\bibfnamefont{J.}~\bibnamefont{Gonzalez-Fraile}},
  \bibnamefont{and} \bibinfo{author}{\bibfnamefont{M.~C.}
  \bibnamefont{Gonzalez-Garcia}} (\bibinfo{year}{2012}), \eprint{1207.1344}.

\bibitem[{\citenamefont{Carmi et~al.}(2012)\citenamefont{Carmi, Falkowski,
  Kuflik, and Volansky}}]{Carmi:2012yp}
\bibinfo{author}{\bibfnamefont{D.}~\bibnamefont{Carmi}},
  \bibinfo{author}{\bibfnamefont{A.}~\bibnamefont{Falkowski}},
  \bibinfo{author}{\bibfnamefont{E.}~\bibnamefont{Kuflik}}, \bibnamefont{and}
  \bibinfo{author}{\bibfnamefont{T.}~\bibnamefont{Volansky}},
  \bibinfo{journal}{JHEP} \textbf{\bibinfo{volume}{1207}}, \bibinfo{pages}{136}
  (\bibinfo{year}{2012}), \eprint{1202.3144}.

\bibitem[{\citenamefont{Weinberg}(1995)}]{citeulike:712984}
\bibinfo{author}{\bibfnamefont{S.}~\bibnamefont{Weinberg}},
  \emph{\bibinfo{title}{The Quantum Theory of Fields (Volume 1)}}
  (\bibinfo{publisher}{Cambridge University Press}, \bibinfo{year}{1995}),
  \bibinfo{edition}{1st} ed., ISBN \bibinfo{isbn}{0521550017},
  \urlprefix\url{http://www.amazon.com/exec/obidos/redirect?tag=citeulike07-20%
\&path=ASIN/0521550017}.

\bibitem[{\citenamefont{Alvarez-Gaum{\'{e}} and
  Ginsparg}(1985)}]{AlvarezGaumé1985439}
\bibinfo{author}{\bibfnamefont{L.}~\bibnamefont{Alvarez-Gaum{\'{e}}}}
  \bibnamefont{and} \bibinfo{author}{\bibfnamefont{P.}~\bibnamefont{Ginsparg}},
  \bibinfo{journal}{Nuclear Physics B} \textbf{\bibinfo{volume}{262}},
  \bibinfo{pages}{439} (\bibinfo{year}{1985}), ISSN \bibinfo{issn}{0550-3213},
  \urlprefix\url{http://www.sciencedirect.com/science/article/pii/055032138590%
3244}.

\bibitem[{\citenamefont{Mambrini}(2011)}]{YannMambrini2011}
\bibinfo{author}{\bibfnamefont{Y.}~\bibnamefont{Mambrini}},
  \bibinfo{journal}{Phys.Rev.} \textbf{\bibinfo{volume}{D84}},
  \bibinfo{pages}{115017} (\bibinfo{year}{2011}), \eprint{1108.0671}.

\bibitem[{\citenamefont{Yaguna}(2009)}]{Yaguna:2008hd}
\bibinfo{author}{\bibfnamefont{C.~E.} \bibnamefont{Yaguna}},
  \bibinfo{journal}{JCAP} \textbf{\bibinfo{volume}{0903}}, \bibinfo{pages}{003}
  (\bibinfo{year}{2009}), \eprint{0810.4267}.

\bibitem[{\citenamefont{Silveira and Zee}(1985)}]{Silveira1985136}
\bibinfo{author}{\bibfnamefont{V.}~\bibnamefont{Silveira}} \bibnamefont{and}
  \bibinfo{author}{\bibfnamefont{A.}~\bibnamefont{Zee}},
  \bibinfo{journal}{Physics Letters B} \textbf{\bibinfo{volume}{161}},
  \bibinfo{pages}{136} (\bibinfo{year}{1985}), ISSN \bibinfo{issn}{0370-2693},
  \urlprefix\url{http://www.sciencedirect.com/science/article/pii/037026938590%
6240}.

\bibitem[{\citenamefont{McDonald}(1994)}]{McDonald:1993ex}
\bibinfo{author}{\bibfnamefont{J.}~\bibnamefont{McDonald}},
  \bibinfo{journal}{Phys.Rev.} \textbf{\bibinfo{volume}{D50}},
  \bibinfo{pages}{3637} (\bibinfo{year}{1994}), \eprint{hep-ph/0702143}.

\bibitem[{\citenamefont{Guo and Wu}(2010)}]{Guo:2010hq}
\bibinfo{author}{\bibfnamefont{W.-L.} \bibnamefont{Guo}} \bibnamefont{and}
  \bibinfo{author}{\bibfnamefont{Y.-L.} \bibnamefont{Wu}},
  \bibinfo{journal}{JHEP} \textbf{\bibinfo{volume}{1010}}, \bibinfo{pages}{083}
  (\bibinfo{year}{2010}), \eprint{1006.2518}.

\bibitem[{\citenamefont{Farina et~al.}(2010)\citenamefont{Farina, Pappadopulo,
  and Strumia}}]{Farina2010329}
\bibinfo{author}{\bibfnamefont{M.}~\bibnamefont{Farina}},
  \bibinfo{author}{\bibfnamefont{D.}~\bibnamefont{Pappadopulo}},
  \bibnamefont{and} \bibinfo{author}{\bibfnamefont{A.}~\bibnamefont{Strumia}},
  \bibinfo{journal}{Physics Letters B} \textbf{\bibinfo{volume}{688}},
  \bibinfo{pages}{329} (\bibinfo{year}{2010}), ISSN \bibinfo{issn}{0370-2693},
  \urlprefix\url{http://www.sciencedirect.com/science/article/pii/S03702693100%
04752}.

\bibitem[{\citenamefont{Djouadi et~al.}(2012)\citenamefont{Djouadi, Lebedev,
  Mambrini, and Quevillon}}]{Djouadi:2011aa}
\bibinfo{author}{\bibfnamefont{A.}~\bibnamefont{Djouadi}},
  \bibinfo{author}{\bibfnamefont{O.}~\bibnamefont{Lebedev}},
  \bibinfo{author}{\bibfnamefont{Y.}~\bibnamefont{Mambrini}}, \bibnamefont{and}
  \bibinfo{author}{\bibfnamefont{J.}~\bibnamefont{Quevillon}},
  \bibinfo{journal}{Phys.Lett.} \textbf{\bibinfo{volume}{B709}},
  \bibinfo{pages}{65} (\bibinfo{year}{2012}), \eprint{1112.3299}.

\bibitem[{\citenamefont{Burgess et~al.}(2001)\citenamefont{Burgess, Pospelov,
  and ter Veldhuis}}]{Burgess:2000yq}
\bibinfo{author}{\bibfnamefont{C.~P.} \bibnamefont{Burgess}},
  \bibinfo{author}{\bibfnamefont{M.}~\bibnamefont{Pospelov}}, \bibnamefont{and}
  \bibinfo{author}{\bibfnamefont{T.}~\bibnamefont{ter Veldhuis}},
  \bibinfo{journal}{Nucl.Phys.} \textbf{\bibinfo{volume}{B619}},
  \bibinfo{pages}{709} (\bibinfo{year}{2001}), \eprint{hep-ph/0011335}.

\bibitem[{\citenamefont{Cerezo et~al.}(2012)\citenamefont{Cerezo, Chala, and
  Lizana}}]{cerezo:2012di}
\bibinfo{author}{\bibfnamefont{R.}~\bibnamefont{Cerezo}},
  \bibinfo{author}{\bibfnamefont{M.}~\bibnamefont{Chala}}, \bibnamefont{and}
  \bibinfo{author}{\bibfnamefont{J.~M.} \bibnamefont{Lizana}},
  \bibinfo{journal}{in preparation}  (\bibinfo{year}{2012}).

\bibitem[{\citenamefont{Chatrchyan et~al.}(2012{\natexlab{a}})}]{:2012sc}
\bibinfo{author}{\bibfnamefont{S.}~\bibnamefont{Chatrchyan}}
  \bibnamefont{et~al.} (\bibinfo{collaboration}{CMS Collaboration})
  (\bibinfo{year}{2012}{\natexlab{a}}), \eprint{1208.0956}.

\bibitem[{\citenamefont{Acosta et~al.}(2003)\citenamefont{Acosta, Affolder, and
  et~all}}]{PhysRevLett.90.081802}
\bibinfo{author}{\bibfnamefont{D.}~\bibnamefont{Acosta}},
  \bibinfo{author}{\bibfnamefont{T.}~\bibnamefont{Affolder}}, \bibnamefont{and}
  \bibinfo{author}{\bibfnamefont{A.}~\bibnamefont{et~all}}
  (\bibinfo{collaboration}{CDF Collaboration}), \bibinfo{journal}{Phys. Rev.
  Lett.} \textbf{\bibinfo{volume}{90}}, \bibinfo{pages}{081802}
  (\bibinfo{year}{2003}),
  \urlprefix\url{http://link.aps.org/doi/10.1103/PhysRevLett.90.081802}.

\bibitem[{\citenamefont{Abazov et~al.}(2008)}]{Abazov:2008vj}
\bibinfo{author}{\bibfnamefont{V.~M.} \bibnamefont{Abazov}}
  \bibnamefont{et~al.} (\bibinfo{collaboration}{D0 Collaboration}),
  \bibinfo{journal}{Phys.Rev.Lett.} \textbf{\bibinfo{volume}{100}},
  \bibinfo{pages}{211803} (\bibinfo{year}{2008}), \eprint{0803.3256}.

\bibitem[{\citenamefont{Beringer et~al.}(2012)}]{Beringer:1900zz}
\bibinfo{author}{\bibfnamefont{J.}~\bibnamefont{Beringer}} \bibnamefont{et~al.}
  (\bibinfo{collaboration}{Particle Data Group}), \bibinfo{journal}{Phys.Rev.}
  \textbf{\bibinfo{volume}{D86}}, \bibinfo{pages}{010001}
  (\bibinfo{year}{2012}).

\bibitem[{\citenamefont{Chatrchyan
  et~al.}(2012{\natexlab{b}})}]{Chatrchyan:2012meb}
\bibinfo{author}{\bibfnamefont{S.}~\bibnamefont{Chatrchyan}}
  \bibnamefont{et~al.} (\bibinfo{collaboration}{CMS Collaboration}),
  \bibinfo{journal}{JHEP} \textbf{\bibinfo{volume}{1208}}, \bibinfo{pages}{023}
  (\bibinfo{year}{2012}{\natexlab{b}}), \eprint{1204.4764}.

\bibitem[{\citenamefont{Chatrchyan
  et~al.}(2012{\natexlab{c}})}]{Chatrchyan:2012it}
\bibinfo{author}{\bibfnamefont{S.}~\bibnamefont{Chatrchyan}}
  \bibnamefont{et~al.} (\bibinfo{collaboration}{CMS Collaboration}),
  \bibinfo{journal}{Phys.Lett.} \textbf{\bibinfo{volume}{B714}},
  \bibinfo{pages}{158} (\bibinfo{year}{2012}{\natexlab{c}}),
  \eprint{1206.1849}.

\bibitem[{\citenamefont{Aad et~al.}(2012{\natexlab{a}})}]{:2012hf}
\bibinfo{author}{\bibfnamefont{G.}~\bibnamefont{Aad}} \bibnamefont{et~al.}
  (\bibinfo{collaboration}{ATLAS Collaboration})
  (\bibinfo{year}{2012}{\natexlab{a}}), \eprint{1209.2535}.

\bibitem[{\citenamefont{Aad et~al.}(2012{\natexlab{b}})}]{:2012dm}
\bibinfo{author}{\bibfnamefont{G.}~\bibnamefont{Aad}} \bibnamefont{et~al.}
  (\bibinfo{collaboration}{ATLAS Collaboration})
  (\bibinfo{year}{2012}{\natexlab{b}}), \eprint{1209.4446}.

\bibitem[{\citenamefont{Chatrchyan et~al.}(2011)}]{Chatrchyan:2011ns}
\bibinfo{author}{\bibfnamefont{S.}~\bibnamefont{Chatrchyan}}
  \bibnamefont{et~al.} (\bibinfo{collaboration}{CMS Collaboration}),
  \bibinfo{journal}{Phys.Lett.} \textbf{\bibinfo{volume}{B704}},
  \bibinfo{pages}{123} (\bibinfo{year}{2011}), \eprint{1107.4771}.

\bibitem[{\citenamefont{Aad et~al.}(2012{\natexlab{c}})}]{Aad:2011fq}
\bibinfo{author}{\bibfnamefont{G.}~\bibnamefont{Aad}} \bibnamefont{et~al.}
  (\bibinfo{collaboration}{ATLAS Collaboration}), \bibinfo{journal}{Phys.Lett.}
  \textbf{\bibinfo{volume}{B708}}, \bibinfo{pages}{37}
  (\bibinfo{year}{2012}{\natexlab{c}}), \eprint{1108.6311}.

\bibitem[{\citenamefont{Barcelo
  et~al.}(2012{\natexlab{b}})\citenamefont{Barcelo, Carmona, Masip, and
  Santiago}}]{Barcelo:2011vk}
\bibinfo{author}{\bibfnamefont{R.}~\bibnamefont{Barcelo}},
  \bibinfo{author}{\bibfnamefont{A.}~\bibnamefont{Carmona}},
  \bibinfo{author}{\bibfnamefont{M.}~\bibnamefont{Masip}}, \bibnamefont{and}
  \bibinfo{author}{\bibfnamefont{J.}~\bibnamefont{Santiago}},
  \bibinfo{journal}{Phys.Lett.} \textbf{\bibinfo{volume}{B707}},
  \bibinfo{pages}{88} (\bibinfo{year}{2012}{\natexlab{b}}), \eprint{1106.4054}.

\bibitem[{\citenamefont{Cagil and Dag}(2012)}]{Cagil:2012py}
\bibinfo{author}{\bibfnamefont{A.}~\bibnamefont{Cagil}} \bibnamefont{and}
  \bibinfo{author}{\bibfnamefont{H.}~\bibnamefont{Dag}} (\bibinfo{year}{2012}),
  \eprint{1203.2232}.

\bibitem[{\citenamefont{Huitu et~al.}(2001)\citenamefont{Huitu, Laitinen,
  Maalampi, and Romanenko}}]{Huitu:2000ut}
\bibinfo{author}{\bibfnamefont{K.}~\bibnamefont{Huitu}},
  \bibinfo{author}{\bibfnamefont{J.}~\bibnamefont{Laitinen}},
  \bibinfo{author}{\bibfnamefont{J.}~\bibnamefont{Maalampi}}, \bibnamefont{and}
  \bibinfo{author}{\bibfnamefont{N.}~\bibnamefont{Romanenko}},
  \bibinfo{journal}{Nucl.Phys.} \textbf{\bibinfo{volume}{B598}},
  \bibinfo{pages}{13} (\bibinfo{year}{2001}), \eprint{hep-ph/0006261}.

\bibitem[{\citenamefont{Akeroyd and Sugiyama}(2011)}]{Akeroyd:2011zza}
\bibinfo{author}{\bibfnamefont{A.~G.} \bibnamefont{Akeroyd}} \bibnamefont{and}
  \bibinfo{author}{\bibfnamefont{H.}~\bibnamefont{Sugiyama}},
  \bibinfo{journal}{Phys.Rev.} \textbf{\bibinfo{volume}{D84}},
  \bibinfo{pages}{035010} (\bibinfo{year}{2011}), \eprint{1105.2209}.

\end{thebibliography}

\end{document}